\documentclass[twocolumn,tighten]{aastex63}
\DeclareUnicodeCharacter{0301}{*************************************}
\usepackage{amsmath}
\usepackage{booktabs,array,dcolumn}
\usepackage{tabularx}
\usepackage{scrextend} 
\usepackage{graphicx,subfigure}
\usepackage{booktabs,array,dcolumn}
\usepackage{tabularx}
\def\msun{$\rm M_{\sun}$}

\newcommand{\hi}{\mbox{\sc{Hi}$\,$}}
\newcommand{\nhi}{$N_{\rm HI}\,$}
\newcommand{\mhi}{$M_{\rm HI}\,$}
\newcommand{\hal}[1]{\ensuremath{{\rm H}\alpha}}

\defcitealias{s19}{S19}

\accepted{\today}
\submitjournal{ApJ}

\begin{document}

\shorttitle{GBT-MHONGOOSE}
\shortauthors{Sardone {\em et al.}}

\title{A Census of the Extended Neutral Hydrogen Around 18 MHONGOOSE Galaxies}

\correspondingauthor{Amy Sardone; NSF Astronomy and Astrophysics Postdoctoral Fellow}
\email{sardone.4@osu.edu}

\newcommand{\OSU}{\affil{Department of Astronomy, The Ohio State University, 140 West 18th Avenue, Columbus, OH 43210, USA}}
\newcommand{\CCAPP}{\affil{Center for Cosmology and Astroparticle Physics, 191 West Woodruff Avenue, Columbus, OH 43210, USA}}
\newcommand{\WVU}{\affil{Department of Physics and Astronomy, West Virginia University, Morgantown, WV 26506}}
\newcommand{\GWAC}{\affil{Gravitational Wave and Cosmology Center, Chestnut Ridge Research Building, Morgantown, WV 26505 USA}}
\newcommand{\GBO}{\affil{Adjunct Astronomer at Green Bank Observatory, Green Bank, WV}}
\newcommand{\ANU}{\affil{Research School of Astronomy \& Astrophysics, Australian National University, Canberra, ACT 2611, Australia}}
\newcommand{\UCT}{\affil{Department of Astronomy, University of Cape Town, Private Bag X3, Rondebosch 7701, South Africa}}
\newcommand{\OAU}{\affil{Laboratoire de Physique et de Chimie de l’Environnement, Observatoire d’Astrophysique de l’Universit\'{e} Ouaga I Pr Joseph Ki-Zerbo (ODAUO), 03 BP 7021, Ouaga 03, Burkina Faso}}
\newcommand{\ASTRON}{\affil{Netherlands Institute for Radio Astronomy (ASTRON), Oude Hoogeveensedijk 4, 7991 PD Dwingeloo, the Netherlands}}
\newcommand{\KAI}{\affil{Kapteyn Astronomical Institute, University of Groningen, PO Box 800, 9700 AV Groningen, The Netherlands}}

\author[0000-0002-5783-145X]{Amy Sardone}
\OSU \CCAPP

\author{D.J. Pisano}
\WVU \GWAC \GBO

\author[0000-0001-9504-7386]{N. M. Pingel}
\ANU \WVU \GWAC

\author[0000-0002-5233-8260]{A. Sorgho}
\UCT

\author[0000-0001-9089-6151]{Claude Carignan}
\UCT \OAU

\author[0000-0001-8957-4518]{W. J. G. de Blok}
\ASTRON \UCT \KAI

\begin{abstract}
\label{sec:chapter3:abstract}

We present the analysis of the diffuse, low column density \hi environment of 18 MHONGOOSE galaxies. We obtained deep observations with the Robert C. Byrd Green Bank Telescope, and reached down to a $3\sigma$ column density detection limit of \nhi $= 6.3 \times 10^{17} \, \rm cm^{-2}$ over a $20 \, \rm km \, s^{-1}$ linewidth. We analyze the environment around these galaxies, with a focus on \hi gas that reaches column densities below \nhi $= 10^{19} \, \rm cm^{-2}$. We calculate the total amount of \hi gas in and around the galaxies revealing that nearly all of these galaxies contained excess \hi outside of their disks. We quantify the amount of diffuse gas in the maps of each galaxy, defined by \hi gas with column densities below $10^{19} \, \rm cm^{-2}$, and find a large spread in percentages of diffuse gas. However, by binning the percentage of diffuse \hi into quarters, we find that the bin with the largest number of galaxies is the lowest quartile ($0-25\%$ diffuse \hi). We identified several galaxies which may be undergoing gas accretion onto the galaxy disk using multiple methods of analysis, including azimuthally averaging column densities beyond the disk, and identifying structure within our integrated intensity (Moment 0) maps. We measured \hi mass outside the disks of most of our galaxies, with rising cumulative flux even at large radii. We also find a strong correlation between the fraction of diffuse gas in a galaxy and its baryonic mass, and test this correlation using both Spearman and Pearson correlation coefficients. We see evidence of a dark matter halo mass threshold of $M_{halo} \sim 10^{11.1}$ \msun{} in which galaxies with high fractions of diffuse \hi all reside below. It is in this regime in which cold-mode accretion should dominate. Finally, we suggest a rotation velocity of $v_{rot} \sim 80 \, \rm km \, s^{-1}$ as an upper threshold to find diffuse gas-dominated galaxies.

\end{abstract}


\section{Introduction} \label{sec:chapter3:intro}
The question of how galaxies get their gas and how they use this gas to continue to form stars remains among the main unanswered questions in astronomy. We know that as the star formation rate density over cosmic time has decreased since a redshift of z$\sim 2$ \citep{2014ARA&A..52..415M}, the neutral hydrogen gas density has remained nearly constant since z$\sim 4$ \citep{2018MNRAS.473.1879R}. \cite{2014A&A...567A..68D} show that mergers of dwarf galaxies with spiral galaxies do not provide sufficient extra mass to sustain star formation in those galaxies. Even so, if star formation in spiral galaxies is not sustained by mergers alone, then they must be accreting the gas from the intergalactic medium (IGM). 

Gas accreting from the IGM flows into the galaxy via diffuse filamentary structures \citep{2003MNRAS.345..349B, 2003ASSL..281..185K, 2005MNRAS.363....2K}. The gas travels from the cosmic web \citep{1996Natur.380..603B} through these filaments, and into the circumgalactic medium (CGM) of galaxies before falling onto the galaxy disk. The process of accretion described here typically proceeds via the hot or the cold mode \citep{2005MNRAS.363....2K}, where gas entering the halo either becomes shock heated to the virial temperature before cooling and condensing onto the galaxy disk or it flows along cold filaments through the CGM and remains cool as is falls onto the disk. While gas in both scenarios is predominantly ionized, the gas accreting via the cold-mode maintains a small neutral fraction, making detection feasible. Cold accretion dominates in the lower mass galaxy regime and favors low galaxy density environments. Observations of the CGM and disks of galaxies encompassing a range of these parameters will provide insight into cold accretion from the IGM and potentially direct detections of this gas falling onto galaxies.

Deep \hi surveys provide the observational link to direct detections of the effects of cold mode accretion from the IGM. The MeerKAT \hi Observations of Nearby Galactic Objects; Observing Southern Emitters (MHONGOOSE\footnote{\url{https://mhongoose.astron.nl}\label{note1}}; \citealt{2016mks..confE...7D}) survey will provide high spatial resolution, high column density sensitivity maps of 30 nearby disk and dwarf galaxies. Most other \hi surveys achieve one or the other, but not both, with the exception of the IMAGINE\footnote{\url{http://www.imagine-survey.org/}\label{note2}} survey with ATCA, which still has a much lower angular resolution than MHONGOOSE. HIPASS \citep{2004AJ....128...16K} and ALFALFA \citep{2011AJ....142..170H} surveys provide an extremely large number of \hi detections, although neither at high resolution nor with column density sensitivities matching ours. High resolution surveys such as THINGS \citep{2008AJ....136.2563W} and HALOGAS \citep{2011A&A...526A.118H} (one of the first high resolution surveys designed to systematically detect accretion of \hi),  achieve column density sensitivities of $\sim 10^{20}\, \rm cm^{-2}$ and $\sim 10^{19}\, \rm cm^{-2}$, respectively. These levels of column densities enable detection of clumpy \hi that could be missed by a single dish telescope, yet would not be low enough to detect any extended or more diffuse gas accreting from the IGM. The MHONGOOSE survey will use the South African MeerKAT radio telescope, a 64-dish precursor to the Square Kilometre Array (SKA), to map these 30 galaxies to a $3\sigma$ column density detection limit of $7.5 \times 10^{18}\, \rm cm^{-2}$ over a $16\, \rm km \, s^{-1}$ linewidth at an angular resolution of $30''$. At their poorest angular resolution, $90''$, this column density limit goes down to $5.5 \times 10^{17}\, \rm cm^{-2}$.  The high resolution maps produced by MHONGOOSE will further our understanding of how galaxies get their gas, how galaxies sustain star formation, and how matter that we can detect relates to the dark matter associated with galaxies, influencing galactic evolution. Specifically, this survey will help us to understand how gas flows in or out of galaxies, the conditions that allow the fuelling of star formation, accretion from the IGM, and ultimately its connection to the cosmic web (e.g. \citealt{2016ASPC..502...55C}). The first of these data are presented in \cite{2020A&A...643A.147D}, detailing their detection of low column density \hi clouds as well as a filament extending off of ESO~302-G014, which could be the result of a minor interaction with a dwarf galaxy.

Early observations of a subset of MHONGOOSE galaxies were presented in \citet[][hereafter S19]{s19}, using MeerKAT, KAT-7, the seven-dish MeerKAT precursor array, and the Robert C. Byrd Green Bank Telescope (GBT). The subset of galaxies mapped with the GBT are presented here, with the description of the MHONGOOSE sample in Section \ref{sec:chapter3:sample}. We describe our GBT observations and data reduction process in Section \ref{sec:chapter3:obs}. Two galaxies not presented in \citetalias{s19}, NGC~1744 and NGC~7424, are presented in Section \ref{sec:chapter3:twosources}. The analysis we performed on all of these galaxies is described in Section \ref{sec:chapter3:analysis}. Our results are presented in Section \ref{sec:chapter3:results}. We discuss these results in Section \ref{sec:chapter3:discussion} and summarize our findings in Section \ref{sec:chapter3:summary}.

\section{MHONGOOSE Galaxy Sample}
\label{sec:chapter3:sample}
The MHONGOOSE sample of galaxies was chosen from galaxies having been previously detected in multiple wavelengths (\hi from HIPASS, \hal{}, optical, infrared, and ultraviolet from the SINGG and SUNGG \citep{2006ApJS..165..307M} surveys). MHONGOOSE galaxies were primarily selected to cover a large range of \hi masses. A secondary parameter used in the selection, the star formation rate (SFR), enables the separation of \hi coincident with the star forming disk from extraplanar \hi (e.g. \cite{2019A&A...631A..50M}). An equal number of galaxies were selected for each of six mass bins within a range of $6 < \rm log_{10}(M_{HI}) < 11$ \msun\ for a total of 30 galaxies with edge-on, face-on, and intermediate inclinations, and a wide range of morphologies from dwarf irregulars to grand-design spirals. The sample of galaxies chosen for this set of observations coincides with the galaxies from the MHONGOOSE sample ($\delta < 10^\circ$) which can be seen by the GBT, which has a lower limit of declinations $\delta > -46^\circ$. This narrows the sample to 18 galaxies. Our GBT-MHONGOOSE sample sacrifices some of the uniformity of the original sample, while continuing to span the full range in \hi masses. These galaxies similarly span a wide range of stellar masses falling between $6.7 < \rm log_{10}(M_{\star}) < 10.7$ \msun\ \citep{2019ApJS..244...24L}. GBT integrated intensity (moment 0) maps and global \hi profiles for 16 of these sources were presented in \citetalias{s19}, with two others, NGC~1744 and NGC~7424, having been observed separately and introduced here. 

\begin{figure}
\centering
\includegraphics[trim=2.5cm 1cm 3cm 1cm, clip, width=\columnwidth]{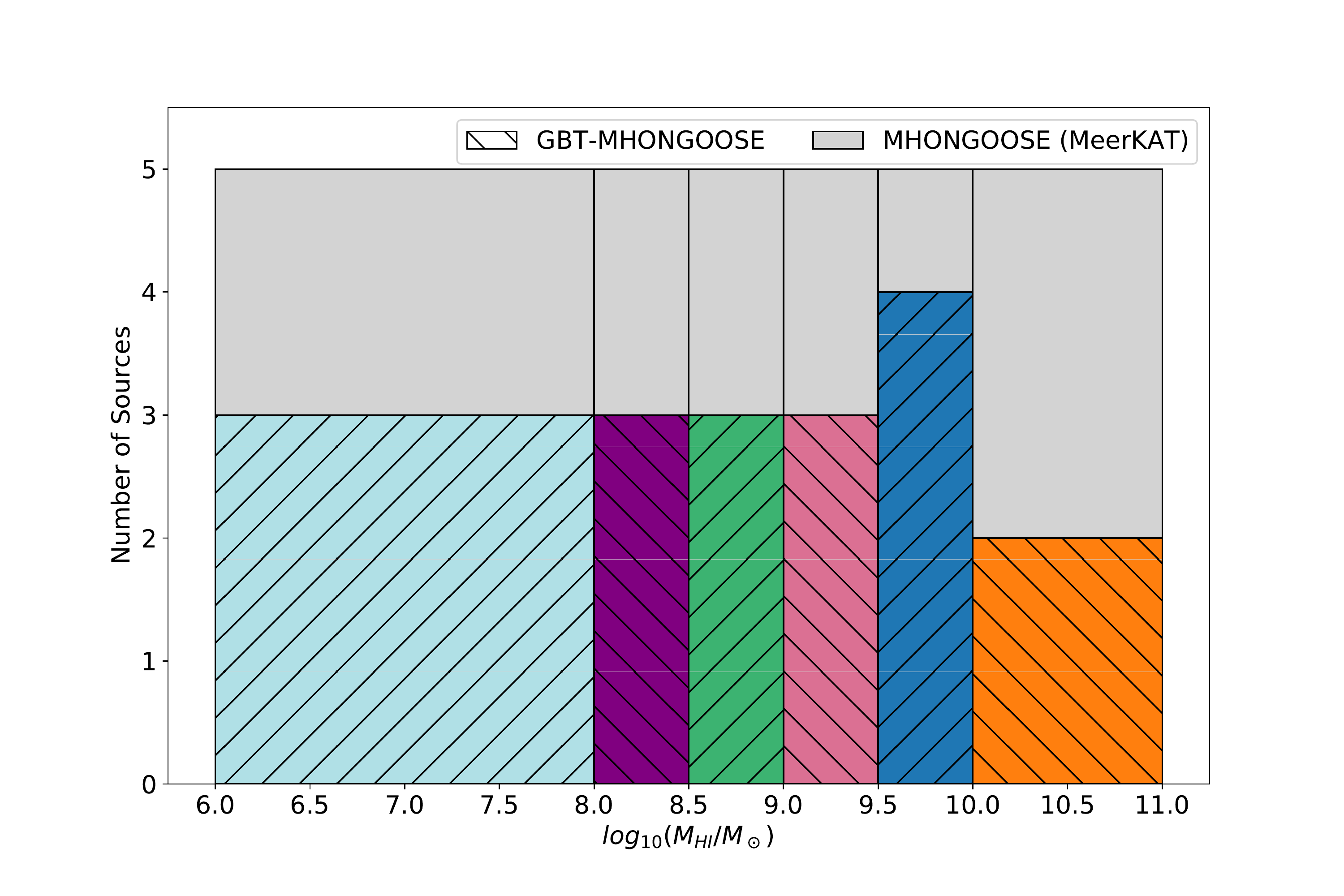}
\caption{Histogram of \hi masses in the MHONGOOSE sample. Each bin is separated by color and hatch marking. The hatch markings (both directions) identify the GBT-MHONGOOSE sample and the gray solid represents the full MHONGOOSE with MeerKAT sample. This histogram demonstrates the coverage of the \hi mass range in the both MHONGOOSE samples. Note that the lowest and highest mass bins are larger due to the scarcity of galaxies meeting the survey criteria in those mass ranges.}
\label{fig:chapter3:histo}
\vspace{0.5cm}
\end{figure}

Each galaxy in the sample presented here, as well as the full MHONGOOSE sample, has a Galactic latitude of $|b| > 30^\circ$, peak \hi fluxes, as detected in HIPASS, of greater than 50 mJy, and Galactic standard of rest velocities $> 200 \, \rm km \, s^{-1}$. Additionally, each source is below a declination of $\delta < 10^\circ$ and is within a distance of 30 Mpc. The MHONGOOSE sample was chosen to uniformly cover the range of \mhi{} listed above, and our GBT sample covers that range nearly uniformly as well. Coverage of the \hi mass range in each bin can be seen in Figure \ref{fig:chapter3:histo}. We note that both the lowest and highest \hi mass bins cover a larger mass range than the others due to the scarcity of galaxies in those mass ranges which also meet the survey criteria. Further details on the MHONGOOSE sample selection can be found in \cite{2016mks..confE...7D}.

\section{Observations and Data Reduction}
\label{sec:chapter3:obs}
Observations for the 16 galaxies presented in \citetalias{s19} were carried out between August 2016 and January 2017 for project GBT16B-212. Each of these galaxies were observed for 10 hours each, a total of 160 hours, with a theoretical brightness temperature noise sensitivity of $\sim 13$ mK over $5.2\,\rm km \, s^{-1}$ channels. This noise sensitivity corresponds to a column density sensitivity of \nhi $\sim 10^{18} \, \rm cm^{-2}$ over the same channel width. The two additional galaxies presented here (NGC~1744 and NGC~7424) were observed over the course of three GBT observing semesters (GBT15B-346, GBT16B-408, and GBT17A-478) from 2015 to 2017, for a total of 20 hours on NGC~1744 and 14 hours on NGC~7424. We observed these two galaxies over a bandwidth of 23.4 MHz, with a frequency resolution of 0.715 kHz. We used the sources 3C48 and 3C147, which have stable, well-understood fluxes, over the course of 12 observing nights to calibrate our data. We smoothed the data using a boxcar function with a final smoothed velocity resolution of $5.1 \, \rm km \, s^{-1}$, or 24.3 kHz. The remainder of the data reduction followed the same approach presented in \citetalias{s19}. 

For each of our 18 galaxies, we used the GBT's VErsatile GBT Astronomical Spectrometer (VEGAS) backend in L-band (1.15-1.73 GHz), where the FWHM beamwidth is $9.1'$, to map $2^\circ \times \, 2^\circ$ regions around each source. Data cubes were made for all 18 sources with pixel sizes of $1.75'$. The 16 sources from \citetalias{s19} were smoothed to a velocity resolution of $6.4 \, \rm km \, s^{-1}$.

We reached rms noise levels of 6 to 20 mK (3.2 - 10.4 mJy) in the cubes, corresponding to a $1\sigma$ column density sensitivity of \nhi $= 1.02 \times 10^{17} \, \rm cm^{-2}$ per $5.1 \, \rm km \, s^{-1}$ channel. At the noise level reached in this cube, a detectable signal at the $5\sigma$ level over a $20 \, \rm km \, s^{-1}$ linewidth is \nhi $= 2 \times 10^{18} \, \rm cm^{-2}$. Our $1\sigma$ mass sensitivities extend down to $1.6 \times 10^5$ \msun.

\section{Properties of galaxies}
\label{sec:chapter3:twosources}
Data cubes and integrated intensity (Moment 0) maps for the 16 sources mentioned above were inspected and searched for anomalous \hi in \citetalias{s19}. In this section we will present the two additional sources, NGC~1744 and NGC~7424, and discuss their properties. These properties are listed in Table \ref{tab:chapter3:derived}. 

\begin{figure*}
\centering
\minipage{0.5\textwidth}
\includegraphics[trim=1.5cm 6cm 2.5cm 6cm, clip,height=8.75cm,width=\linewidth]{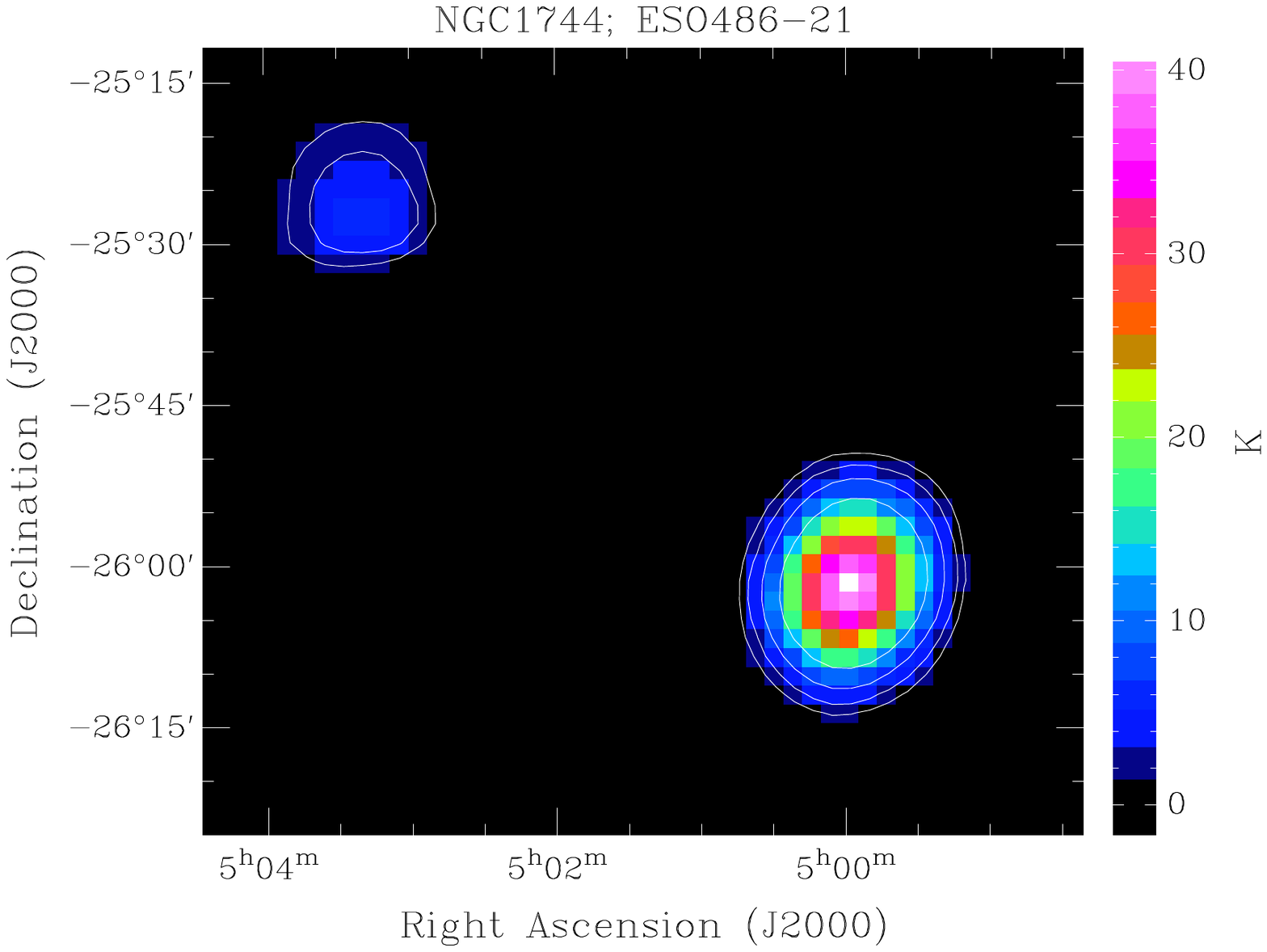}
\endminipage\hfill
\minipage{0.5\textwidth}
\includegraphics[trim=1cm 6cm 2cm 6cm, clip,width=\linewidth]{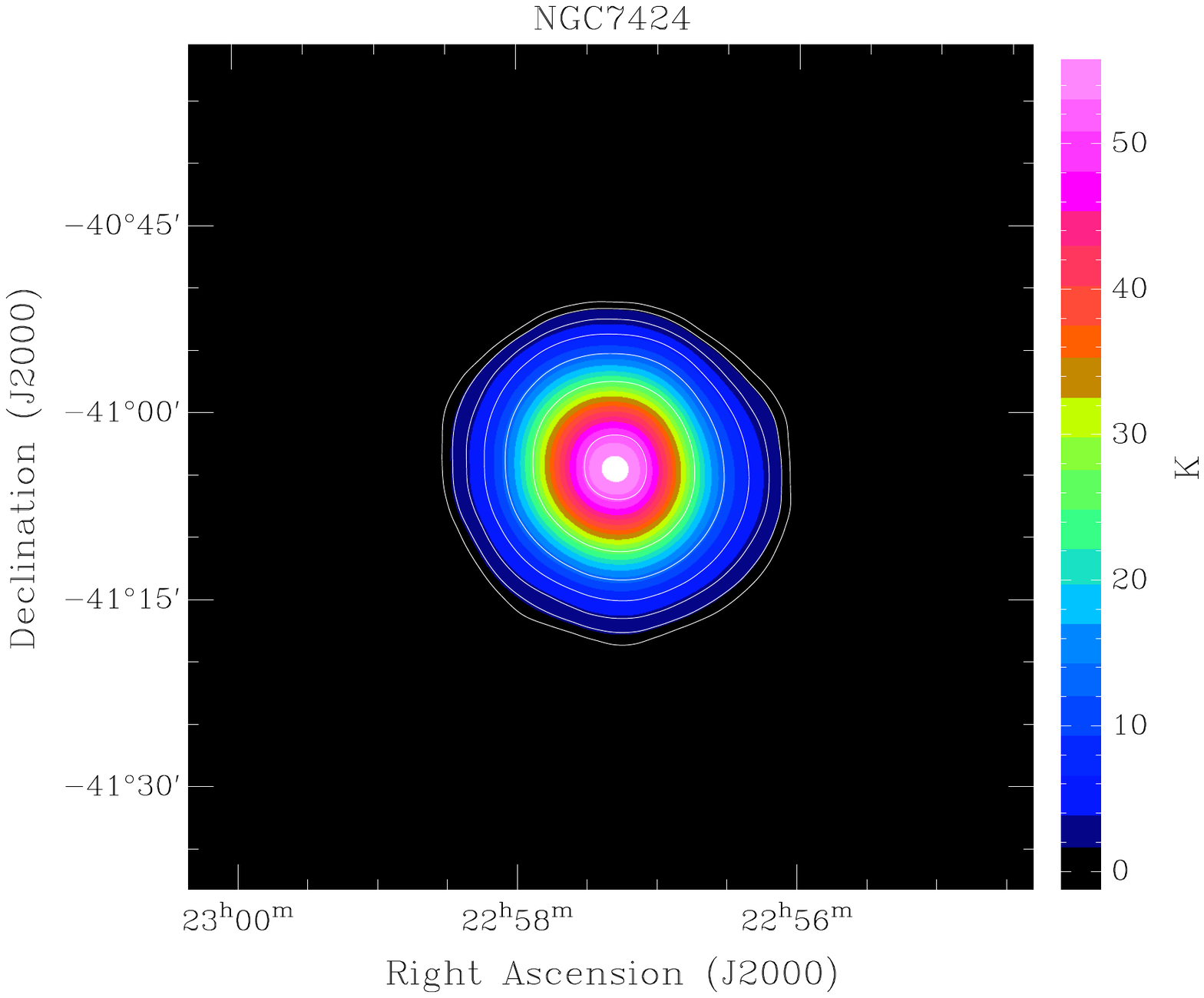}
\endminipage\hfill
\caption{{\it Left.} Integrated intensity (moment 0) map of NGC~1744 with companion, ESO486-21 in upper left. Contours at 10, 20, and 40 times the integrated rms noise of 0.15 K. The column density equivalent to these contour levels is \nhi = 0.69, 1.3, 2.7, and $5.5 \times 10^{19} \, \rm cm^{-2}$. {\it Right.} Integrated intensity (moment 0) map of NGC~7424. Contours at 5, 10, 20, 40, 80, and 160 times the integrated rms noise of 0.16 K. The column density equivalent to these contour levels is \nhi = 0.7, 1.48, 2.9, 5.9, and $11.9 \times 10^{19} \, \rm cm^{-2}$.
\label{fig:chapter3:NGC1744mom0}}
\vspace{0.5cm}
\end{figure*}

\label{fig:chapter3:moment-maps}

\subsection{NGC~1744}
NGC~1744 is an inclined ($69.9^\circ$) SBcd\footnote{\url{http://leda.univ-lyon1.fr}\label{note3}} galaxy. It is one of the more massive galaxies in our sample, which we measured the total \hi mass to be $7.6\times 10^{9}$ \msun\ ($\pm 2.1\times 10^{8}$). We measured a total integrated flux of 174.7 Jy km/s with an rms noise of $\sigma_{rms} = 3.2$ mJy. We measured the full width at 20\% maximum of the velocity profile to be 207.2 km/s with a systemic velocity of 743 km/s, which we used to calculate a dynamical mass (see subsection \ref{derivedprop}, equation \ref{Mdyneq}) of NGC~1744 of $3.6 \times 10^{10}$ \msun. These and other properties of NGC~1744 were tabulated in Table \ref{tab:chapter3:derived} and discussed further in Section \ref{sec:chapter3:analysis}.

\begin{figure*}
\centering
\minipage{0.45\textwidth}
    \includegraphics[trim=0.5cm 0cm 2cm 0cm, clip,width=\linewidth]{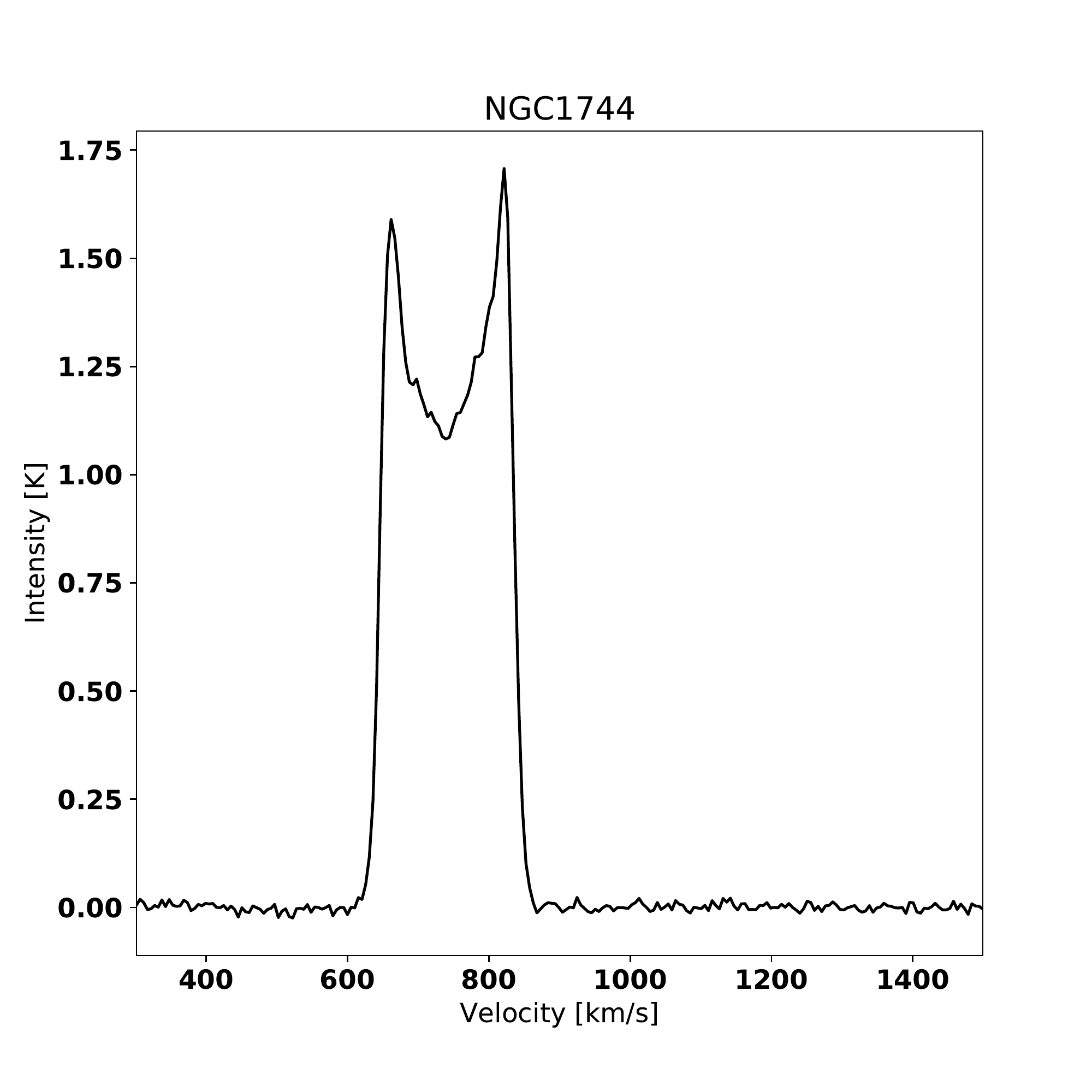}
\endminipage
\minipage{0.45\textwidth}
    \includegraphics[trim=0.5cm 0cm 2cm 0cm, clip,width=\linewidth]{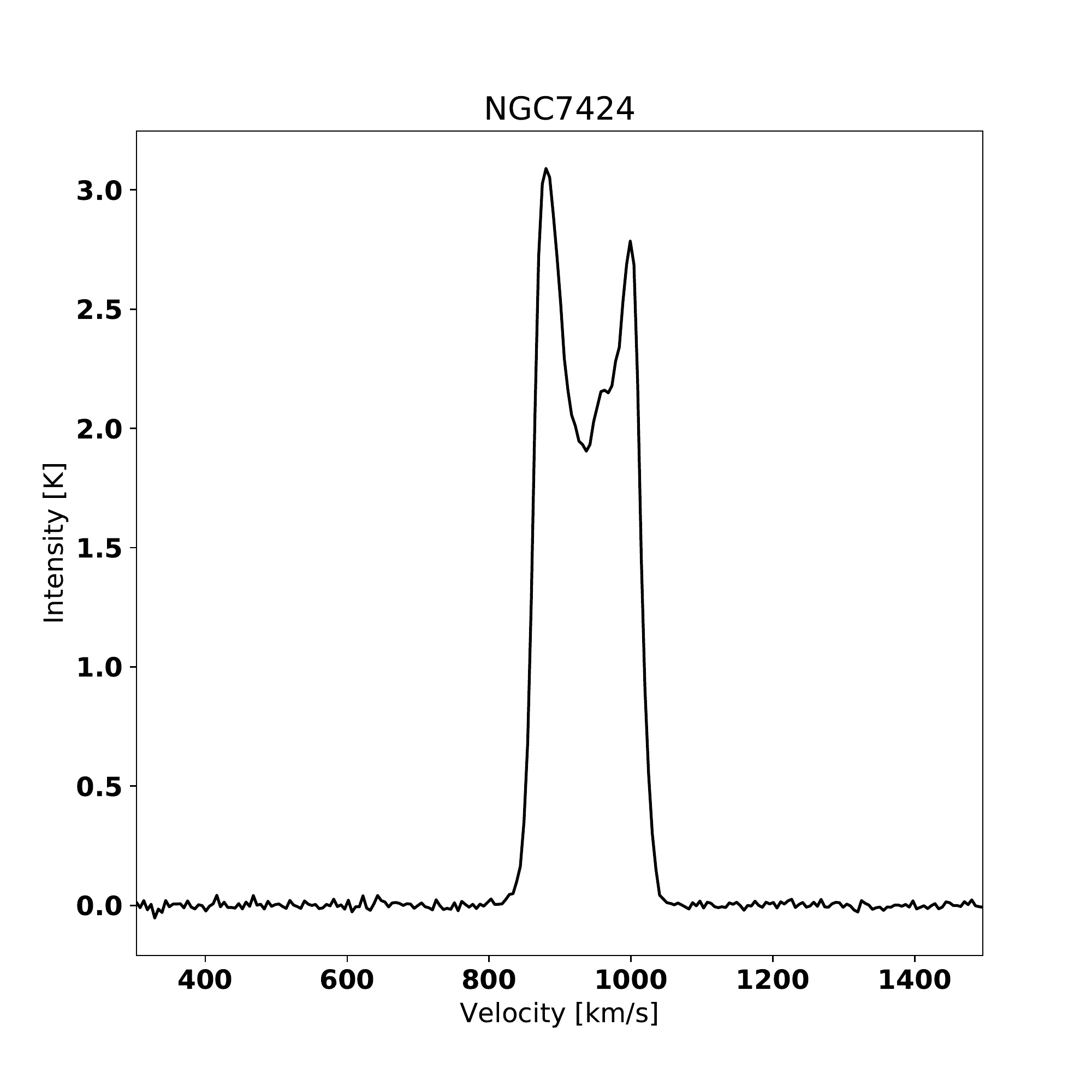}
\endminipage\hfill
\minipage{0.33\textwidth}
    \includegraphics[trim=0.5cm 1cm 2cm 1cm, clip,width=\linewidth]{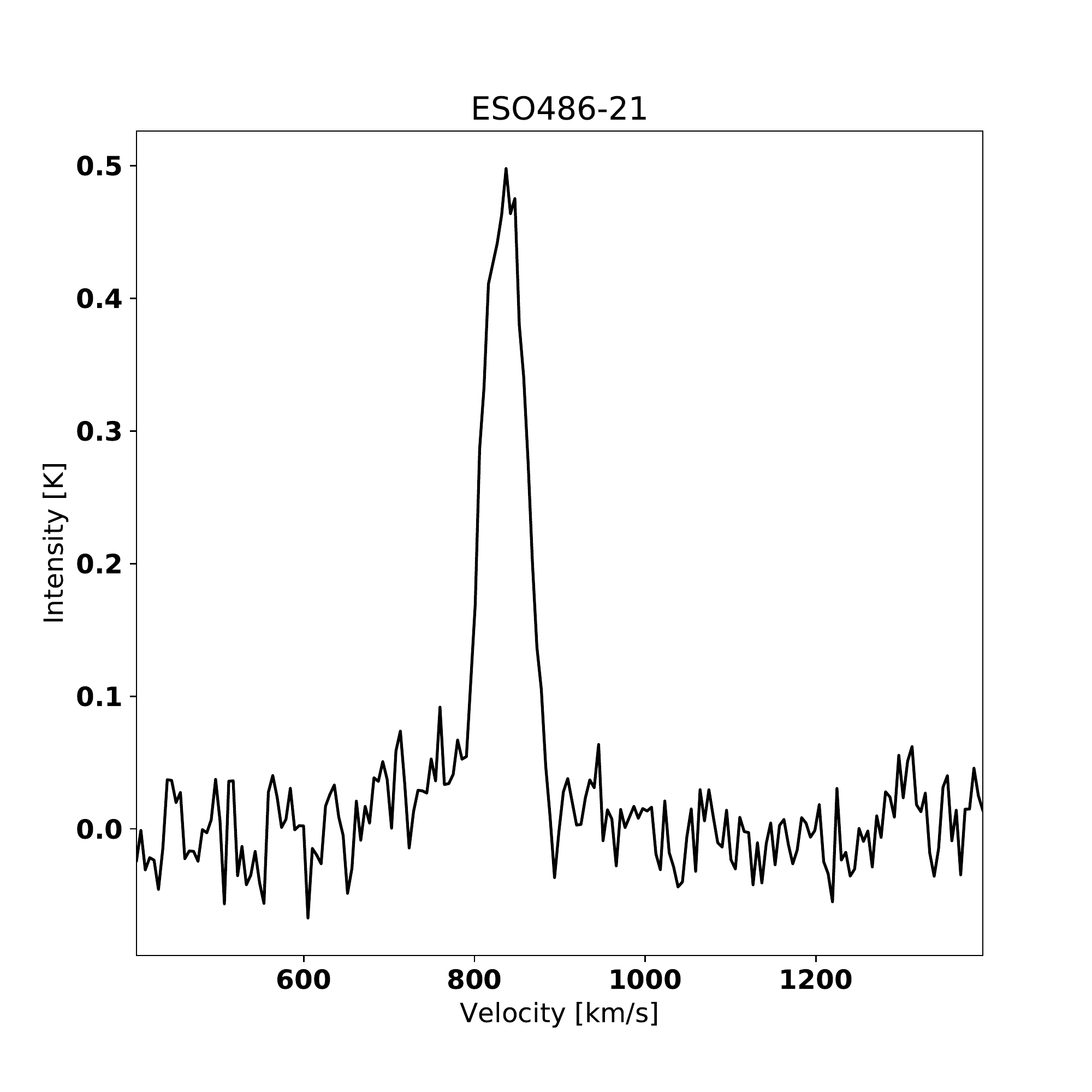}
\endminipage\hfill
\minipage{0.33\textwidth}
    \includegraphics[trim=0.6cm 1cm 2cm 1cm, clip,width=\linewidth]{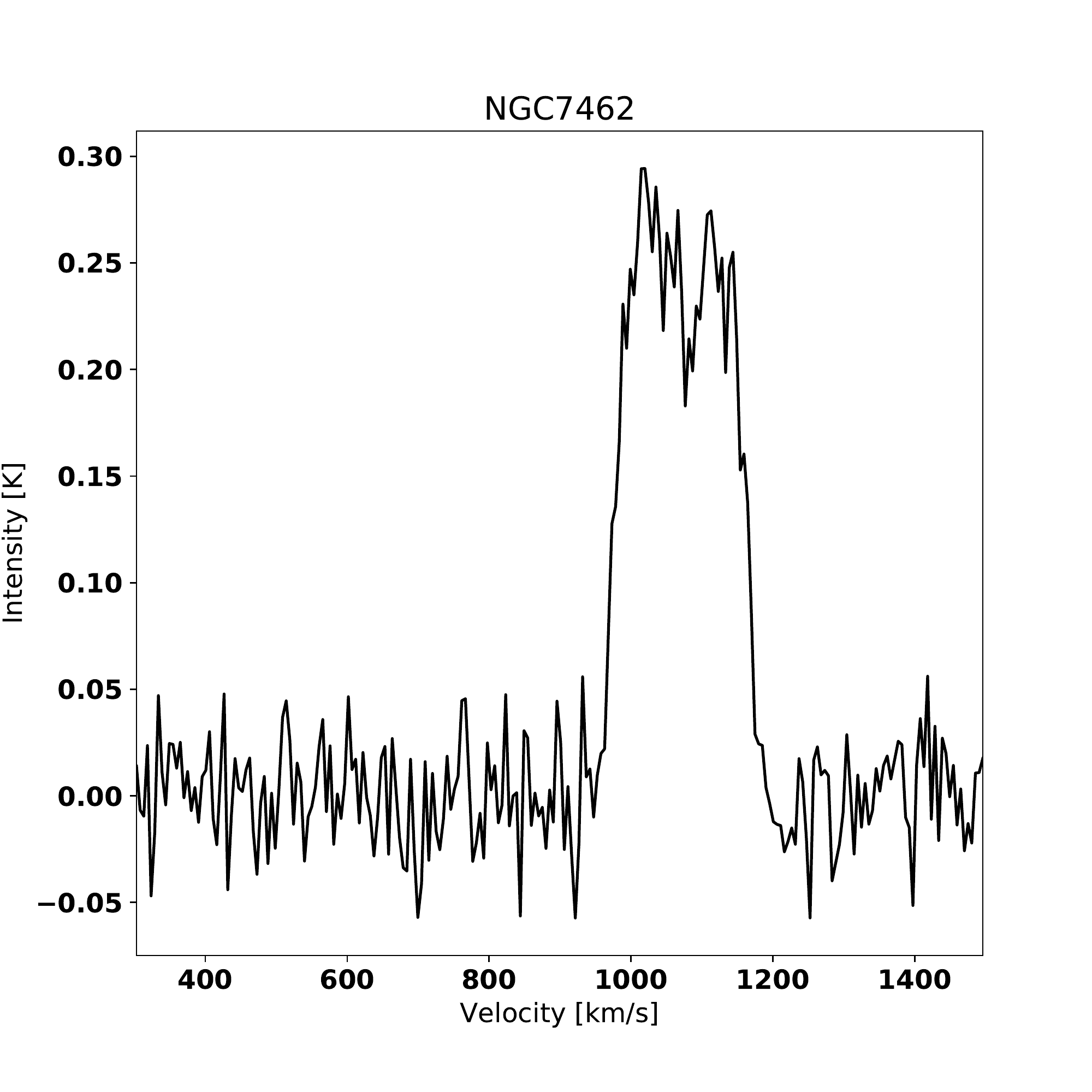}
\endminipage\hfill
\minipage{0.33\textwidth}
    \includegraphics[trim=0.6cm 1cm 2cm 1cm, clip,width=\linewidth]{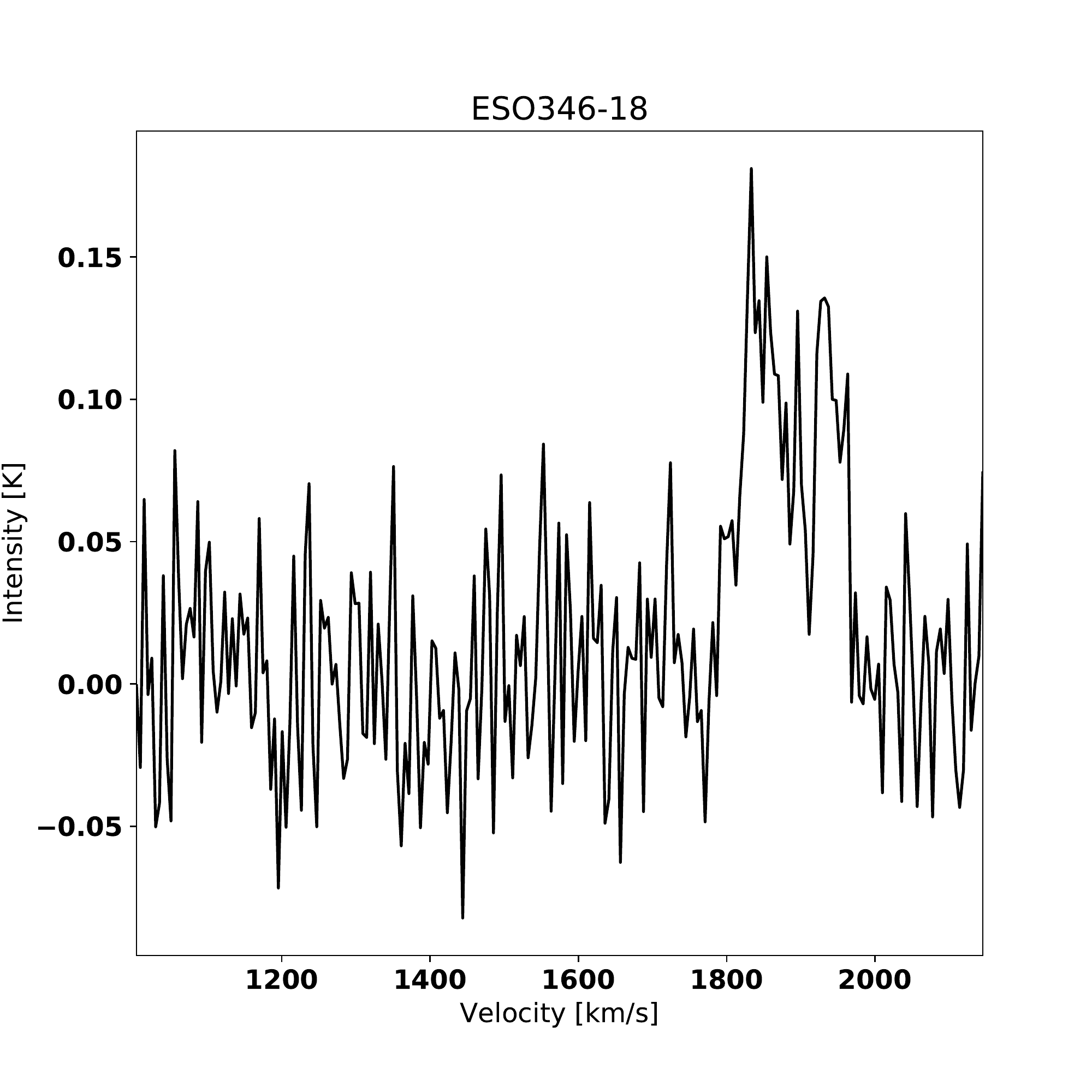}
\endminipage\hfill
\caption{Total integrated \hi{} profiles. Top row: NGC~1744 (left) and NGC~7424 (right). Bottom row: ESO486-21 detected in the NGC~1744 cube (left), and detected in the NGC~7424 cube is NGC7462  (center) and ESO346-18 (right).
\label{fig:chapter3:N1744-7424}}
\vspace{0.5cm}
\end{figure*}

We detected one additional source in the NGC~1744 data cube. Sbc\footref{note1} galaxy, ESO~486-G021 (Figure \ref{fig:chapter3:N1744-7424}), is detected with an integrated \hi flux density of 24.5 Jy $\rm km \, s^{-1}$. The systemic velocities measured from the two galaxies differ by $\sim 100 \, \rm km \, s^{-1}$, and while this is low enough to investigate potential interaction, we find their physical separation to be too great to make this likely. The angular separation is $58'$, and at a distance of 12 Mpc, that becomes a physical separation of $\sim 200$ kpc. We calculated the Jacobi radius, the maximum radius expected of a central galaxy and companion galaxy system, with

\begin{equation}
    r_{\rm J} = R_{Sep} \left(\frac{M_{Comp}}{3M_{Central}}\right)^{1/3} \, .
\end{equation}

\noindent Here, $R_{Sep}$ is the physical separation between the centers of the two galaxies, $M_{Comp}$ is the dynamical mass of the companion galaxy, and $M_{Central}$ is the dynamical mass of the central galaxy. This gives us a maximum expected radius of $\sim 79$ kpc using the dynamical mass, derived from the \hi rotation velocity, of the companion of $2.2 \times 10^9$ \msun. It is evident that the 200 kpc projected separation is too large to consider plausible interaction. Further, we do not detect any extraplanar or anomalous \hi around NGC~1744 at these levels. A total \hi intensity map including both NGC~1744 and ESO~486-G021 can be seen in Figure \ref{fig:chapter3:NGC1744mom0}. Spectra of NGC~1744 can be seen in Figure \ref{fig:chapter3:N1744-7424}. At a mass sensitivity in this cube of $7.2\times 10^{5}$ \msun, and because we did not detect any anomalous \hi, we can say that apart from ESO~486-G021 there are no additional sources with $3\sigma$ \hi mass greater than $7.2\times 10^{5}$ \msun.

\subsection{NGC~7424}
NGC~7424 is a mostly face-on SABcd galaxy \citep{1991rc3..book.....D}, and one of the most \hi massive sources in the survey. It falls into the mass bin: $10 < \rm log_{10}(M_{HI}) < 11$ \msun, one of the more unexplored mass regimes in low column density \hi studies. We measured a total \hi mass of $2.4\times 10^{10}$ \msun\ ($\pm 6.2\times 10^{8}$) in NGC~7424, where the mass sensitivity in the cube is $2.3\times 10^{6}$ \msun\ corresponding to a $1\sigma$ column density sensitivity of \nhi $= 9.7 \times 10^{16} \, \rm cm^{-2}$ per $5.1 \, \rm km \, s^{-1}$channel. NGC~7424's global \hi profile and moment map can be seen in Figure \ref{fig:chapter3:N1744-7424}.

We detected two additional galaxies within our map of NGC~7424: the Sd galaxy ESO~346-G018 with an angular separation of $58'$, and the SBbc galaxy NGC~7462 with an angular separation from NGC~7424 of $64'$. However, due to their positions at the edges of our cube, we can not confidently measure the total \hi. Instead, we can provide the following lower limits. We measured an integrated flux of 8.8 Jy km/s for ESO~346-G018, with a linewidth at 20\% maximum of 176 km/s around a systemic velocity of 1875 km/s. NGC~7462 has an integrated flux lower limit of 25.0 Jy km/s, a 20\% maximum linewidth of 205 km/s, and a systemic velocity of 1066 km/s. The total integrated profiles were created for each galaxy and can be seen in Figure \ref{fig:chapter3:N1744-7424}. No other significant detections of \hi were seen throughout the cube. 

\section{Analysis}
\label{sec:chapter3:analysis}
Each source in this survey was analyzed by extracting information from the data cubes and deriving individual galaxy properties, followed by creating both masked and unmasked total integrated intensity (moment 0) maps, and measuring the global properties of the galaxy environments from these images. Moment 0 maps for each source can be found in \citetalias{s19}. In doing this, we can identify the statistical properties over a wide population of galaxies.

\subsection{Total integrated \texorpdfstring{\hi}{hi} flux profiles}
We created total integrated \hi flux profiles for each galaxy using the \texttt{MIRIAD} task \textsc{mbspect}. \textsc{mbspect} makes a measurement of the data cube over a wide velocity range, where channels including emission are masked, and a first-order polynomial is fitted to the baseline, removing residual baseline variation. We obtain the $1\sigma$ rms noise over emission-free channel ranges, as well as the total integrated flux, and the linewidths of each source at 20\% (W20) and 50\% (W50) of the peak flux value. These measured values are listed in Table \ref{tab:chapter3:derived}. Each galaxy's global line profile can be seen in Figures \ref{fig:chapter3:ESO300-G016} through \ref{fig:chapter3:UGCA250}.

\subsection{Derived galaxy properties}
\label{derivedprop}
For each data cube, we calculate the per channel $1\sigma$ column density sensitivity. We use measurements from \textsc{mbspect} to derive physical properties of each galaxy in the sample, such as the total \hi masses, total dynamical masses, and neutral gas fractions.

Column densities are calculated with 
\begin{equation}
    N_{\rm HI} = 1.822 \times 10^{18} \, \left(\frac{T_{\rm B}}{\rm K}\right) \, \left(\frac{dv}{\rm km \, s^{-1}}\right) \rm cm^{-2},
\end{equation}

\noindent where $T_{\rm B}$ is the brightness temperature in units of Kelvin, and $dv$ is the resolution of our data in $\rm km \, s^{-1}$. When estimating the detectable column density level in the cube, we use a $5\sigma$ limit and a minimum linewidth of a typical \hi detection ($20 \, \rm km \, s^{-1}$).

We integrate over the total flux values to estimate the \hi mass of each source, assuming the \hi is optically thin, using: 
\begin{equation}
    M_{\hi} = 2.36 \times 10^5 \, D^2 \int_{v_1}^{v_2} S(v) dv \, \rm M_{\sun} .
\end{equation}

\noindent Here, the distance, $D$, is in Mpc, and $\int_{v_1}^{v_2}S(v) dv$ is the total integrated flux  over velocities enclosing the \hi profile $v_1$ to $v_2$, and is in units of Jy km $\rm s^{-1}$. The brightness temperatures from each cube were divided by the $1.86 \, \rm K \, Jy^{-1}$ gain of the GBT to obtain units of Jy. We calculated this gain using an aperture efficiency of $\sim0.65$ \citep{2011A&A...536A..81B} for the GBT at 1420 MHz. 

In order to calculate the total dynamical mass of each galaxy, we first calculated the physical \hi diameter, $D_{\hi}$, using the \cite{1997A&A...324..877B} scaling relation (Eq. 2.5), which is dependent on the optical diameter of the galaxy, $D_{25}$, measured at the $\rm 25^{th} \, mag \, arcsec^{-2}$ isophote in the B-band. We used values for $D_{25}$ from the Lyon-Meudon Extragalactic Database (LEDA). We use the dynamical mass equation:
\begin{equation}
\label{Mdyneq}
    M_{\rm dyn} = 2.3 \times 10^5 \, \left(\frac{v_{rot}/sin(i)}{\rm km \, s^{-1}}\right)^2 \left(\frac{r}{\rm kpc}\right) \rm M_{\sun} ,
\end{equation}
\noindent where $v_{\rm rot}$ is the rotation velocity in $\rm km \, s^{-1}$, and $i$ the inclination, also taken from LEDA. These properties obtained from NED and LEDA can be seen in Table \ref{tab:chapter:looked-up}. We use our linewidth at 20\% maximum to estimate $v_{\rm rot}$ as $v_{\rm rot} = W_{20}/2$. Dynamical masses are calculated inside the \hi radius, $r$ in kpc, calculated as $D_{\hi}/2$.

The last property we derive using the information at hand, is the neutral \hi fraction. This is defined as the fraction of \hi mass in the galaxy and is calculated with $f_{\hi} = M_{\rm HI}\, / \, M_{\rm dyn}$. Each of these derived properties are listed in Table \ref{tab:chapter3:derived} for every galaxy in the sample.

\begin{deluxetable}{lcccccr}
\tabletypesize{\footnotesize}
\tablecaption{Sample of MHONGOOSE Galaxies}
\setlength{\tabcolsep}{1pt} 
\renewcommand{\arraystretch}{1.25}
\tablewidth{0pt}
\tablehead{
\colhead{Source} &
\colhead{R.A.(J2000)} &
\colhead{Decl.(J2000)} &
\colhead{D } &
\colhead{incl. } &
\colhead{$D_{25}$} &
\colhead{$\rm log_{10}M_{\star}$} \\
\colhead{} &
\colhead{[h:m:s]} &
\colhead{[$^{\circ}$:$'$:$''$]} &
\colhead{[Mpc]} &
\colhead{[deg]} &
\colhead{$'$} &
\colhead{[\msun]}
}

\decimalcolnumbers
\startdata
ESO300G014 & 03:09:37 & -41:01:50 & 12.9 & 61.2 & 4.47 & 8.72 \\
ESO300G016 & 03:10:10 & -40:00:11 & 9.3 & 35.6 & 0.78 & ... \\
ESO302G014 & 03:51:40 & -38:27:08 & 11.7 & 27.6 & 1.35 & 7.79 \\
ESO357-G007 & 03:10:24 & -33:09:22 & 17.8 & 72.0 & 1.29 & 8.31 \\
KK98-195 & 13:21:08 & -31:31:45 & 5.2 & 55.7 & 0.45 & ... \\
KKS2000-23 & 11:06:12 & -14:24:26 & 12.7 & 90.0 & 0.6 & ... \\
NGC1371 & 03:35:01 & -24:56:00 & 20.4 & 47.5 & 4.9 & 10.7 \\
NGC1592 & 04:29:40 & -27:24:31 & 13.0 & 64.4 & 1.02 & 8.13 \\
NGC1744 & 04:59:57 & -26:01:20 & 10.0 & 69.9 & 5.25 & 9.19 \\
NGC3511 & 11:03:23 & -23:05:12 & 14.2 & 72.6 & 6.03 & 9.63 \\
NGC5068 & 13:18:54 & -21:02:21 & 6.9 & 30.1 & 7.41 & 9.32 \\
NGC5170 & 13:29:48 & -17:57:59 & 28.0 & 90.0 & 7.94 & 10.72 \\
NGC5253 & 13:39:55 & -31:38:24 & 3.0 & 70.1 & 5.01 & 8.57 \\
NGC7424 & 22:57:18 & -41:04:14 & 13.5 & 32.4 & 5.01 & 9.31 \\
UGCA015 & 00:49:49 & -21:00:54 & 3.3 & 67.4 & 1.62 & 6.74 \\
UGCA250 & 11:53:24 & -28:33:11 & 24.4 & 90.0 & 3.63 & 9.74 \\
UGCA307 & 12:53:57 & -12:06:21 & 8.6 & 62.0 & 1.82 & 7.96 \\
UGCA320 & 13:03:16 & -17:25:23 & 7.7 & 90.0 & 6.76 & 8.11 \\
\enddata
\tablecomments{Positions listed in Columns 2-3 and distances in 4 gathered from NED. Inclination angles from face-on in columns 5 and the optical diameter, $D_{25}$, measured at the $\rm 25^{th} \, mag \, arcsec^{-2}$ isophote in the B-band values taken from HyperLEDA: http://leda.univ-lyon1.fr. Stellar masses were obtained from \cite{2019ApJS..244...24L}.}
\label{tab:chapter:looked-up}
\end{deluxetable}

\subsection{GBT beam model}
\label{GBTbeam-model}
One of the main science goals of this survey is to detect low column density \hi around our sample of galaxies. In order to distinguish between low column density \hi from an extragalactic source and the low level radiation entering the sidelobes of the GBT's main beam, we use a beam model which measures these sidelobe levels precisely, modeling the beam response of an unresolved point source. We adopt the beam model used by \cite{2018ApJ...865...36P}. Without this model, emission entering the nearest sidelobes from the main beam could be confused for low column density emission from our sources. For each source in our sample, we used the GBT beam model as a template to regrid each data cube to have $4''$ pixels with size $1024 \times 1024$ using the \texttt{MIRIAD} tasks \textsc{imgen} and \textsc{regrid}. 

\subsection{Integrated intensity images (Moment 0)}
We created integrated intensity images, or Moment 0 images, for each source in our sample. An unmasked image was made by integrating pixel values over a channel range in the data cube containing emission from the central source. We also created masked Moment 0 images with the intention of separating signal from noise when searching for low column density \hi. In order to do this, we set to zero the pixels with values below three times the noise in that cube, and again integrate over the same channel range. After integration over the channel range we selected, we expect the $\sigma_{rms}$ noise in each map (masked and unmasked) to increase by a factor of: 
\begin{equation}
    \sigma_N = \sqrt{N} \, \sigma_{rms}
\end{equation}

\noindent where $N$ is the number of channels integrated over, and $\sigma_{rms}$ is the root-mean-squared noise per channel in the data cube. We calculated a $1\sigma$ noise map for both the masked and unmasked images this way. We then create our maps using a $3\sigma$ cutoff in the same way as \cite{2018ApJ...865...36P} by creating a signal-to-noise (S/N) map for both the masked and unmasked images. The S/N map for the unmasked image is made by dividing the unmasked integrated image by the unmasked $1\sigma$ noise map, while the masked S/N map is created by dividing the image where pixels below $3\sigma_{rms}$ were set to zero by the masked $1\sigma$ noise image. From each of these S/N maps, we calculated a $3\sigma$ threshold by taking the mean value of the pixels with S/N values falling between 2.75 and 3.25. We then use this new $3\sigma$ value to mask, or set to zero, the pixels in our masked integrated image that fall below this value. In doing this, we are confident in our characterization of the noise in our maps.

\subsection{Cumulative \texorpdfstring{\hi}{hi} vs. \texorpdfstring{\nhi}{nhi}}
\label{subsec:analysis:cumuHI}
We want to find out how much \hi mass is contained at different column density levels in each galaxy. In doing this, we will be able to identify the amount of low column density \hi in the galaxy as a percentage of the total \hi mass. Accordingly, in the way described in \cite{2018ApJ...865...36P}, we bin the column densities above the $3\sigma$ level in our unmasked image and calculate the percentage of \hi mass that falls into each \nhi bin by converting each pixel in the image to a column density level, and subsequently into an \hi mass. We determine the percentage of \hi mass at or above a given column density bin, and can therefore quantify the percentage of \hi mass below some column density threshold. Each galaxy is normalized to that galaxy's maximum \hi mass in the cumulative \hi vs. \nhi plots in Figures \ref{fig:chapter3:ESO300-G016} through \ref{fig:chapter3:UGCA250}.

In addition to the data from each cube, we use the beam models described above to determine whether each source can be reliably analyzed. Each beam model for each galaxy is scaled to the maximum column density of the associated GBT image from the data and we analyze it in the same way as the GBT images described in the previous paragraph, calculating the cumulative \hi mass in each \nhi bin. In doing this, we can determine if the source is following the response of the GBT beam or deviating from it. If the source follows the beam, we can determine that the source is unresolved, and therefore does not satisfactorily fill the beam, causing the \hi column density to spread out over the beam area appearing as if low column density \hi was detected. To this end, we would omit these unresolved sources from our statistical analysis of the disk. We distinguish which sources are unresolved by looking at the cumulative \hi mass ratio of the data versus the model. If the data follows the beam response of the model, we determine that source to be unresolved. Data points above the model is characterized as resolved, and if the error bars fully clear the beam model, we categorize it as very well resolved. 

We can infer additional information about the amount of low column density \hi in the galaxy by looking at the shape of the profile in relation to the GBT beam model. If a source contains excess diffuse \hi, we will see a positive deviation in the data at low column densities as compared to the beam model. Several of the resolved sources follow the shape of the beam model, and they flatten out at or above the $3\sigma$ column density threshold described in Section \ref{GBTbeam-model}, which is indicative of the scarcity of low column density \hi in that galaxy. If, instead, we see an increase at low column densities as compared to the beam model, then we know we are detecting an excess of low column density \hi within that halo, which could be diffuse gas or clumpy material spread over the beam. This excess could be the result of a number of scenarios including detection of a low column density companion, extended or clumpy \hi clouds, the presence of a $>3\sigma$ noise artefact, or the detection of accretion of low column density \hi from the CGM. For each image showing a higher fraction of low column density \hi, we search for $>3\sigma$ artefacts in order to rule out that possibility, and look for higher S/N regions in both the cubes and the integrated images.

\begin{deluxetable*}{lccccccccccr}
\tabletypesize{\footnotesize}
\tablecaption{\hi Measurements and Derived Properties of MHONGOOSE Galaxies}
\renewcommand{\arraystretch}{1.25}
\tablewidth{0pt}
\tablehead{
\colhead{Source} &
\colhead{$\sigma_{rms}$} &
\colhead{$S_{\hi}$} &
\colhead{$\rm V_{sys}$} &
\colhead{$\rm W_{50}$} &
\colhead{$\rm W_{20}$} &
\colhead{$\rm log_{10}$\nhi$_{1\sigma}$} &
\colhead{$\rm log_{10}$\nhi$_{3\sigma}$} &
\colhead{$\rm log_{10}M_{1\sigma}$} &
\colhead{$\rm log_{10}M_{\hi}$} &
\colhead{$\rm log_{10}M_{dyn}$} &
\colhead{$f_{\hi}$} \\
\colhead{} &
\colhead{[mJy]} &
\colhead{[$\rm Jy\,km\,s^{-1}$]} &
\colhead{[$\rm km\,s^{-1}$]} &
\colhead{[$\rm km\,s^{-1}$]} &
\colhead{[$\rm km\,s^{-1}$]} &
\colhead{[$\rm cm^{-2}$]} &
\colhead{[$\rm cm^{-2}$]} &
\colhead{[\msun]} &
\colhead{[\msun]} &
\colhead{[\msun]} &
}

\decimalcolnumbers
\startdata
ESO300G014 & 10.4 & 33.6 & 955.0 & 129.2 & 145.2 & 17.35 & 18.33 & 6.42 & 9.12 & 10.36 & 0.058 \\
ESO300G016 & 9.1 & 4.9 & 710.5 & 30.1 & 43.9 & 17.29 & 18.27 & 6.07 & 8.0 & 8.78 & 0.167 \\
ESO302G014 & 9.6 & 13.7 & 869.3 & 65.5 & 87.1 & 17.31 & 18.29 & 6.29 & 8.65 & 9.91 & 0.055 \\
ESO357-G007 & 4.9 & 15.0 & 1118.3 & 118.4 & 148.5 & 17.02 & 18.0 & 6.37 & 9.05 & 9.91 & 0.137 \\
KK98-195 & 5.9 & 8.1 & 570.6 & 27.1 & 42.2 & 17.1 & 18.08 & 5.38 & 7.71 & 7.95 & 0.577 \\
KKS2000-23 & 5.3 & 14.2 & 1037.5 & 80.2 & 96.8 & 17.06 & 18.03 & 6.11 & 8.73 & 9.02 & 0.515 \\
NGC1371 & 7.6 & 90.9 & 1454.3 & 386.4 & 403.8 & 17.21 & 18.19 & 6.68 & 9.95 & 11.63 & 0.021 \\
NGC1592 & 4.8 & 5.8 & 942.9 & 55.1 & 96.2 & 17.01 & 17.99 & 6.08 & 8.36 & 9.34 & 0.104 \\
NGC1744 & 3.2 & 174.7 & 743.4 & 191.1 & 207.2 & 16.75 & 17.82 & 5.59 & 9.62 & 10.56 & 0.112 \\
NGC3511 & 8.3 & 74.6 & 1130.8 & 272.4 & 308.7 & 17.25 & 18.23 & 6.4 & 9.55 & 11.11 & 0.028 \\
NGC5068 & 4.8 & 191.5 & 667.7 & 67.9 & 108.0 & 17.01 & 17.99 & 5.53 & 9.33 & 10.53 & 0.063 \\
NGC5170 & 6.2 & 106.2 & 1546.7 & 504.0 & 523.0 & 17.13 & 18.1 & 6.87 & 10.29 & 11.94 & 0.023 \\
NGC5253 & 8.3 & 57.6 & 404.5 & 63.4 & 99.1 & 17.25 & 18.23 & 5.05 & 8.09 & 9.38 & 0.051 \\
NGC7424 & 5.6 & 297.2 & 934.9 & 152.6 & 170.5 & 16.99 & 18.06 & 6.1 & 10.11 & 10.99 & 0.13 \\
UGCA015 & 5.2 & 5.2 & 293.6 & 25.5 & 43.2 & 17.05 & 18.02 & 4.93 & 7.12 & 8.23 & 0.078 \\
UGCA250 & 7.6 & 87.6 & 1700.6 & 272.2 & 289.5 & 17.21 & 18.19 & 6.83 & 10.09 & 11.03 & 0.115 \\
UGCA307 & 5.0 & 32.0 & 822.9 & 68.5 & 95.2 & 17.03 & 18.01 & 5.74 & 8.75 & 9.42 & 0.211 \\
UGCA320 & 7.8 & 133.9 & 749.1 & 106.3 & 126.9 & 17.23 & 18.2 & 5.84 & 9.27 & 10.08 & 0.156 \\
\enddata
\tablecomments{(1) Source name. (2) measured rms noise. (3) Total integrated flux. (4) Systemic velocity. (5) Linewidth at 50\% maximum. (6) Linewidth at 20\% maximum. (7) $1\sigma_{rms}$ column density sensitivity per channel. (8) $3\sigma_{rms}$ column density level over a 20 $\rm km \, s^{-1}$ linewidth. (9) \hi mass sensitivity. (10) \hi mass. (11) Dynamical mass. (12) Neutral gas fraction.}
\label{tab:chapter3:derived}
\end{deluxetable*}

\subsection{Radial \texorpdfstring{\nhi}{nhi}}
We also want to determine how \hi{} column densities (\nhi) behave as a function of distance from the galaxy. The optimal way of identifying abundances in low column density \hi is to take the average of the \hi column density levels in annuli around the galaxy. This allows us to characterize the azimuthally averaged \nhi at various physical galaxy radii. We define the annuli for each galaxy using the masked rather than unmasked images so as to reduce the effect of quantifying low level noise as signal. Each image is regridded to have $1024\times1024$ pixels corresponding to $4''$ per pixel for the purpose of precision when quantities inside each annulus are calculated. The GBT beam's FWHM of $9.1'$ sets a lower limit to the radius of the smallest annulus to be $4.5'$, or $\sim 68$ pixels. The maximum radius of the largest annulus is calculated as the largest multiple of the 68 pixel radius that lies within the image size. For most of our sources, we calculate this to be eight annuli, and for images in which we use a smaller region around the galaxy, we end up with five annuli. The \nhi within each annulus is then averaged by summing the values greater than $3\sigma$ and dividing this by the total number of pixels in the annulus. Dividing by the total, rather than the number of pixels above $3\sigma$, we run the risk of underestimating the average \nhi in the annulus, but we are avoiding a bias toward higher \nhi values which would result from the average using only values $>3\sigma$.

Once again, we want to be able to compare the data against the GBT beam model to look for deviation from the response of the beam. Each model is scaled to the maximum \nhi value in the image, and drops off at larger radii. The galaxies in our sample follow the response of the beam at low radii, and begin to flatten out around the $3\sigma$ threshold. Some of the galaxies flatten out above this threshold, which could indicate a smooth extent of the galaxy rather than an ionization edge at a particular column density level around that galaxy. The $1\sigma$ and $5\sigma$ thresholds are indicated for each galaxy as the dashed horizontal lines in Figures \ref{fig:chapter3:ESO300-G016} through \ref{fig:chapter3:UGCA250}. 

We can also infer details about the environments of these galaxies by looking at the shape of the radial \nhi profile. If the data follows the noise levels out to large radii, we can then say that there is no low column density \hi surrounding that galaxy for levels at or above the rms noise sensitivity reached in that cube. However, several of the galaxies show a positive deviation in the averaged \nhi at these large radii. For each galaxy in which this positive deviation is identified, we investigate the cube and integrated maps in order to identify a source for the excess \hi. The excess could be due to detection of a companion, an extended \hi cloud, a higher $\sigma$ noise spike, or the accretion of \hi onto the galaxy from the CGM. 

\subsection{Radial Flux}
We can trace the flux in each galaxy out to the same physical radii that we explored in the radial \nhi analysis using the same method to calculate each annuli. We use the unmasked images for this analysis as this is the best way to characterize the total flux of the galaxy at various radii, since our unmasked images contain all the flux detected from the source. 

Since we are measuring the cumulative flux at physical radii, we should see the profile continue to increase linearly as the area inside each annulus increases linearly if the average \nhi described above is flat. If the radial flux profile deviates from a positive linear trend, we know to look for an increase in low column density gas around that galaxy. A dip in the profile is likely indicative of an unphysical negative feature in the image, causing the cumulative flux to decrease at that radius. If indeed we see rising cumulative flux levels at increasing radii, these would be consistent with the detection of \hi at large impact parameters as is seen in \cite{2020ApJ...898...15D} where \hi is detected at each point along the minor axis out to $\sim 120$ kpc.

\section{Results}
\label{sec:chapter3:results}
\begin{figure*}
\centering
\includegraphics[trim=0cm 0cm 0cm 0cm, clip,width=\textwidth]{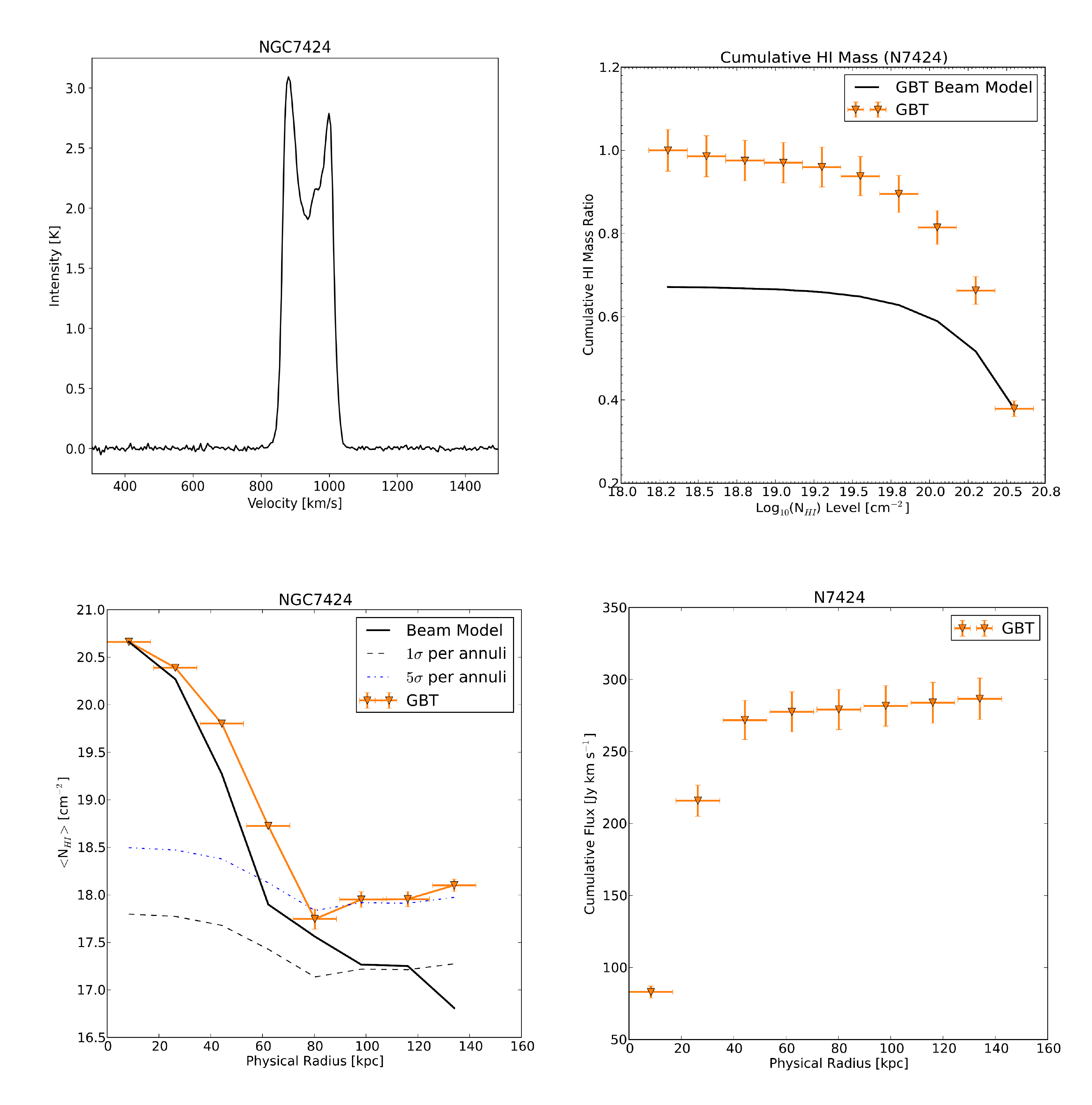}
\caption{{\it Top left}: Total integrated \hi profile. {\it Top right}: Cumulative \hi mass. The total \hi mass in the moment 0 map is plotted by the fraction of gas in column density bins, and compared to the GBT beam model, scaled to the peak column density. {\it Lower left}: Azimuthally averaged \nhi. Column density averaged over annuli extending radially from the center of the galaxy and compared to the GBT beam model. The black dashed line characterizes the $1\sigma$ noise in each annulus, and the blue dot-dash line represents the $5\sigma$ noise in each annulus. {\it Lower right}: Cumulative flux. Flux in each of those annuli are summed to obtain a measure of the total flux out to the edge of each map. 
\label{fig:chapter3:NGC7424_4x4}}
\vspace{0.5cm}
\end{figure*}

Most of the galaxies in our sample are moderately to well-resolved in the GBT beam, and give us reliable statistical properties. As described in Section \ref{subsec:analysis:cumuHI}, a resolved source is defined by a positive offset of the data from the GBT beam model in the cumulative \hi mass vs. \nhi plots and in the radially averaged \nhi vs. physical radius plots. The plots in Figures \ref{fig:chapter3:ESO300-G016} through \ref{fig:chapter3:UGCA250} show, clockwise from top left, the total integrated \hi profile vs. velocity, the cumulative \hi mass vs. \nhi, the cumulative flux vs. physical radius, and the azimuthally averaged \nhi levels vs. physical radius. For reference, we have put this analysis for NGC~7424 in Figure \ref{fig:chapter3:NGC7424_4x4}, while the remainder of the analysis for each galaxy in our sample can be found in the Appendix. The results from analysis of all of our sources will be described below, and they have been grouped by low (Bin 1), intermediate (Bins 2-5), and high (Bin 6) \hi mass displayed in Figure \ref{fig:chapter3:histo}.

\subsection{Low \texorpdfstring{\mhi ($6 < \rm log_{10}(M_{HI}) < 8$ \msun)}{bin1}}
Figure \ref{fig:chapter3:ESO300-G016} displays a very slight positive offset of ESO~300-16 from the GBT model in the cumulative \hi mass vs. column density, where the lowest column density levels deviate in the positive direction. At the smallest radii seen in the radially averaged \nhi plot, we can see that this source is not well resolved until we move beyond the disk, where our underestimated \nhi values are consistent with the noise and begin to rise. The cumulative flux continues to rise even at large distances from the disk.

We see the response of the data in the plots of KK98-195 (Figure \ref{fig:chapter3:KK98-195}) are fairly unremarkable. We see a very slight positive deviation of the cumulative \hi mass as compared to the GBT beam model at the lowest \nhi level, demonstrating a small fraction (less than 10\%) of the \hi mass in the galaxy to be at the lowest column densities. The average \nhi flattens outside the disk in the radially averaged \nhi plot, and the flux increases linearly in the cumulative flux plot.

UGCA015 (Figure \ref{fig:chapter3:UGCA015}) is $\sim 40'$ in angular distance from the large galaxy, NGC~247, which shows up in our data cube at overlapping velocities. For this reason, we analyzed a smaller region around UGCA015 in order to exclude possible emission from NGC~247. We see a positive offset in the \hi mass at the lower column densities. While we were not able to extend out to farther radii, we note that the cumulative flux again does not begin to flatten beyond the disk.

Peak column densities in each of these low \mhi galaxies sit right around \nhi$ = 10^{19} \, \rm cm^{-2}$. None of the three galaxies in this mass range display any positive deviation from the noise level in the azimuthally averaged \nhi plots, but each one shows an increase in the fraction of \hi mass at the lowest column densities. The largest of these fractions comes from UGCA015, where the offset from the beam model deviates significantly, making up over 20\% of the mass fraction from the lowest column densities. Each of the radially averaged cumulative flux profiles continue to rise as far as our maps reach. This indicates that we have not yet reached the end of the \hi emission in our maps of these smallest galaxies.

\subsection{Intermediate \texorpdfstring{\mhi ($8 < \rm log_{10}(M_{HI}) < 10$ \msun)}{bin2}}
\label{subsec:bin2}
Figure \ref{fig:chapter3:ESO302-G014} displays the properties of ESO~302-14, where we see an increase at the lowest column densities in the cumulative \hi mass plot, flattening outside the disk in the radially averaged \nhi plot, and increasing flux levels in the cumulative flux plot. \cite{2020A&A...643A.147D} detect a filament extending off the disk of ESO~302-14 as well as an \hi cloud with a peak column density of \nhi$ = 4 \times 10^{19} \, \rm cm^{-2}$ about 30 kpc south of the center of the galaxy. These structures are significantly smaller than the FWHM beamwidth of the GBT and this emission would be spread out over the beam, diluting the signal. 

NGC~1592 (Figure \ref{fig:chapter3:NGC1592}) has a peak column density level of just over \nhi$ = 10^{19} \, \rm cm^{-2}$. The cumulative flux rises at large radii. 

The properties of NGC~5253 are shown in Figure \ref{fig:chapter3:NGC5253}. There is an increase in the azimuthally averaged column density levels around 25 kpc, which is associated with one bright region in the integrated intensity (moment 0) image. This bright region has no associated optical counterpart, previously identified \hi cloud, nor does the spectra show anything we would consider to be RFI. Thus, we considered it to be a candidate for accretion of \hi onto the galaxy. 

ESO~357-007 (Figure \ref{fig:chapter3:ESO357-G007}) contains increasing amounts of diffuse \hi beginning around \nhi$ = 10^{18.8} \, \rm cm^{-2}$, comprising over 20\% of the cumulative \hi mass fraction. We can see in the azimuthally averaged \nhi plot that the galaxy does not appear to be resolved in the GBT beam at low physical radii of the galaxy, and displays an increasing average \nhi at the radii beyond the disk. The cumulative flux does show a non-linear increase at the largest physical radii mapped. In the absence of evidence of a companion galaxy, or any bright artefact in the data cube causing the cumulative flux to increase, we are led to believe that this evidence of excess neutral hydrogen at large radii could be accretion through the CGM onto ESO~357-007. Alternatively, this excess of low column density \hi could be indicative of a large reservoir of \hi sitting in the halo of ESO~357-007.

KKS2000-23 shows a significant portion of its \hi mass to be made up of low column density gas, as can be seen in the cumulative \hi mass plot in Figure \ref{fig:chapter3:KKS2000-23}. However, inspection of the data cube and the moment map reveal a higher level of noise in the image, making it difficult to determine if the low column density gas comes from real emission or is some artefact from the observation. This particular issue could be alleviated with additional observations to verify the presence or absence of low \nhi gas. 

UGCA307 also shows an increase in low column density gas in Figure \ref{fig:chapter3:UGCA307}. However, the azimuthally averaged \nhi vs. physical radius displays a steep increase between 55 kpc and 65 kpc. We inspected the data cube and discovered a small galaxy, which we believe is LCRS B125208.8-112329, residing 72 kpc away in projected distance. This can also be seen in the cumulative flux vs. physical radius plot, where the flux increases around the same physical distance. This would be the first velocity information available for this galaxy, with a central velocity of 863 km/s.

The spiral galaxy ESO~300-14 can be seen in Figure \ref{fig:chapter3:ESO300-G014}. The total integrated flux of ESO~300-14 displays the expected double horn feature for an inclined, rotating galaxy. It is resolved in the GBT beam, where the cumulative flux follows the shape of the GBT beam model, and is positively offset in relation to that model. The azimuthally averaged column densities vs. physical radius show a rising \nhi level outside the disk, and the cumulative flux at increasing physical radii increases linearly.

Similarly, the face-on galaxy, NGC~5068, exhibits similar behavior when comparing these same properties. The cumulative \hi mass plot in Figure \ref{fig:chapter3:NGC5068} shows that $\sim 99 \%$ of the mass in the galaxy is seen in column densities above \nhi $= 10^{19} \rm \, cm^{-2}$. This is substantiated in both the average \nhi and cumulative flux vs. physical radii plots, where there is no evidence of an increase in \hi gas at large radii.

We detected two additional sources in the region close to UGCA320: UGCA319 at $\sim 41$ kpc, and a second detection around $\sim 61$ kpc projected distances. The closer companion, UGCA319 can be seen as a slight bump in the averaged \nhi around $\sim45$ kpc in Figure \ref{fig:chapter3:UGCA320}. The second \hi detection lacks a known counterpart in \hi, but may be associated with the small galaxy LEDA 886203, the only optical galaxy near this position. These two companions are likely what is causing the gradual increase in cumulative \hi flux at radii larger than 40 kpc. This would also be the first redshift information available on LEDA 886203 with a central velocity of 727 km/s. 

NGC~1371 is well-resolved in the GBT beam, as seen in the cumulative \hi mass plot in Figure \ref{fig:chapter3:NGC1371}. In most of these plots, NGC~1371 behaves as expected, with the exception of an artefact at $\sim 150$ kpc as seen in the radially averaged \nhi plot. 

One of our more massive galaxies, NGC~1744, shows no signs of excess amounts of low column density gas as seen in the cumulative \hi mass plot in Figure \ref{fig:chapter3:NGC1744}. We can see in the radially averaged \nhi plot that this source is well-resolved, and shows an unusually high average \nhi at radii beyond the disk. This could be indicative of smooth accretion of \hi, a reservoir of \hi halo gas, or a higher noise floor than we had estimated. 

NGC~3511 (Figure \ref{fig:chapter3:NGC3511}) shows large deviation at low column density levels, where the cumulative amount of \hi increases almost linearly through \nhi$ = 10^{18.6} \, \rm cm^{-2}$. It is resolved in the GBT beam as seen in both the cumulative mass plot and the averaged \nhi plot. Similar to NGC~1744, this source displays higher \nhi values out into the halo.

NGC~5170, in contrast with NGC~1744 which resides in the same mass bin, does show an increase towards lower column density \hi mass. Figure \ref{fig:chapter3:NGC5170} reveals a sharp change in the slope of the cumulative \hi mass at column densities lower than $\sim 10^{19.4} \, \rm cm^{-2}$. The azimuthally averaged \nhi plot does show an increase in average \nhi around 175 kpc. Inspection of the moment 0 image reveals some bright, filamentary structure at the same physical distance, which could be accreting \hi gas. Without any indication of a nearby optical counterpart associated with those bright regions which appear to lack optical counterparts, we consider this a candidate for accretion from the CGM.

\subsection{High \texorpdfstring{\mhi ($10 < \rm log_{10}(M_{HI}) < 11$ \msun)}{bin6}}
We can see NGC~7424, one of the most \hi massive galaxies in our sample, in Figure \ref{fig:chapter3:NGC7424_4x4}. The data in the averaged \nhi vs. physical radius plot shows slightly higher and gradually increasing values of \nhi at radii outside the disk. Visual inspection of the image did not reveal any regions of particular interest. 

One of our largest sources, UGCA250 (Figure \ref{fig:chapter3:UGCA250}), occupies our highest mass bin: $10 < \rm log_{10}(M_{HI}) < 11$ \msun. The cumulative \hi mass increases at low column densities, and the cumulative flux continues to grow linearly as the radius increases. Inspection of the data cube revealed a bright, positive stripe across the top of UGCA250, at the same declination as another bright source on the edge of our cube, UGCA247. Within that stripe is a possible new \hi detection which is coincident with the only catalogued galaxy within the size of the GBT beam at this position, FLASH J115508.00-282045.1 at a radius of $\sim 185$ kpc. 

\section{Discussion}
\label{sec:chapter3:discussion}

\subsection{\hi in the CGM}
Our analysis of the cumulative \hi mass allowed us to quantify the total \hi mass in the moment maps. Our maps were made at a constant angular size of $2^{\circ} \times 2^{\circ}$, so we do not have a consistent physical region around each mapped source where initial map sizes range from 105 -- 977 kpc on one side. The measurement of the \hi mass is made in the moment map where we can subtract off the \hi mass of the disk, which is derived from the integrated spectra of each galaxy. The total \hi mass in the moment map less the integrated \hi disk mass from the spectra gives us some value for the amount of mass outside the disk. We refer to this amount as $f_{CGM}$. This fraction represents the total amount of \hi mass in the CGM in relation to the \hi disk mass: $f_{CGM} = \rm M_{CGM} / M_{disk}$. Out of our 18 galaxies, 16 contained an excess amount of gas outside their disk by 0.02 to 3 times the amount of \hi in the disk. The only two galaxies we did not detect \hi in their CGM were NGC~1744 and NGC~7424. We can see this in both the cumulative \hi mass plots and the cumulative flux plots for both of these galaxies. In both of these galaxies, the cumulative \hi mass stays mostly flat at column densities below its turnover (\nhi $\lesssim 10^{19.5}\, \rm cm^{-2}$) and the cumulative flux does not rise at large distance from the galaxy. This same flattening also occurs with NGC~5068, which contains only 2\% more \hi outside its disk. This tells us that there is very little, if any, low column density gas outside the disks of these three galaxies. We generally find that maps covering a smaller physical area contain a smaller total amount of \mhi outside of the disk, and larger physical maps contain a larger total amount of \mhi outside of the disk. This may seem like an intuitive result if \hi\ permeates the CGM of most galaxies, and yet it has evaded most observational studies of \hi emission in the CGM of galaxies. This result is consistent with \cite{2020ApJ...898...15D} where \hi emission was detected in each measurement at increasing distances along the minor axis of the two galaxies, NGC~891 and NGC~4565. These detections could be the outcome of ubiquitous CGM \hi which, in mapping observations such as this current work, would result in a larger amount of \hi detected over larger physical areas. A study of this type could be improved on with a consistent physical parameter guiding the size of the area observed, such as the virial radius of a galaxy. Mapping the full extent of the halo, as the upcoming Parkes-IMAGINE survey (Sardone et al., in prep) will do, would give insight into the amount of \hi\ permeating the entire CGM of a galaxy.

\begin{figure}
\centering
\includegraphics[trim=0cm 0cm 0cm 0cm, clip, width=\columnwidth]{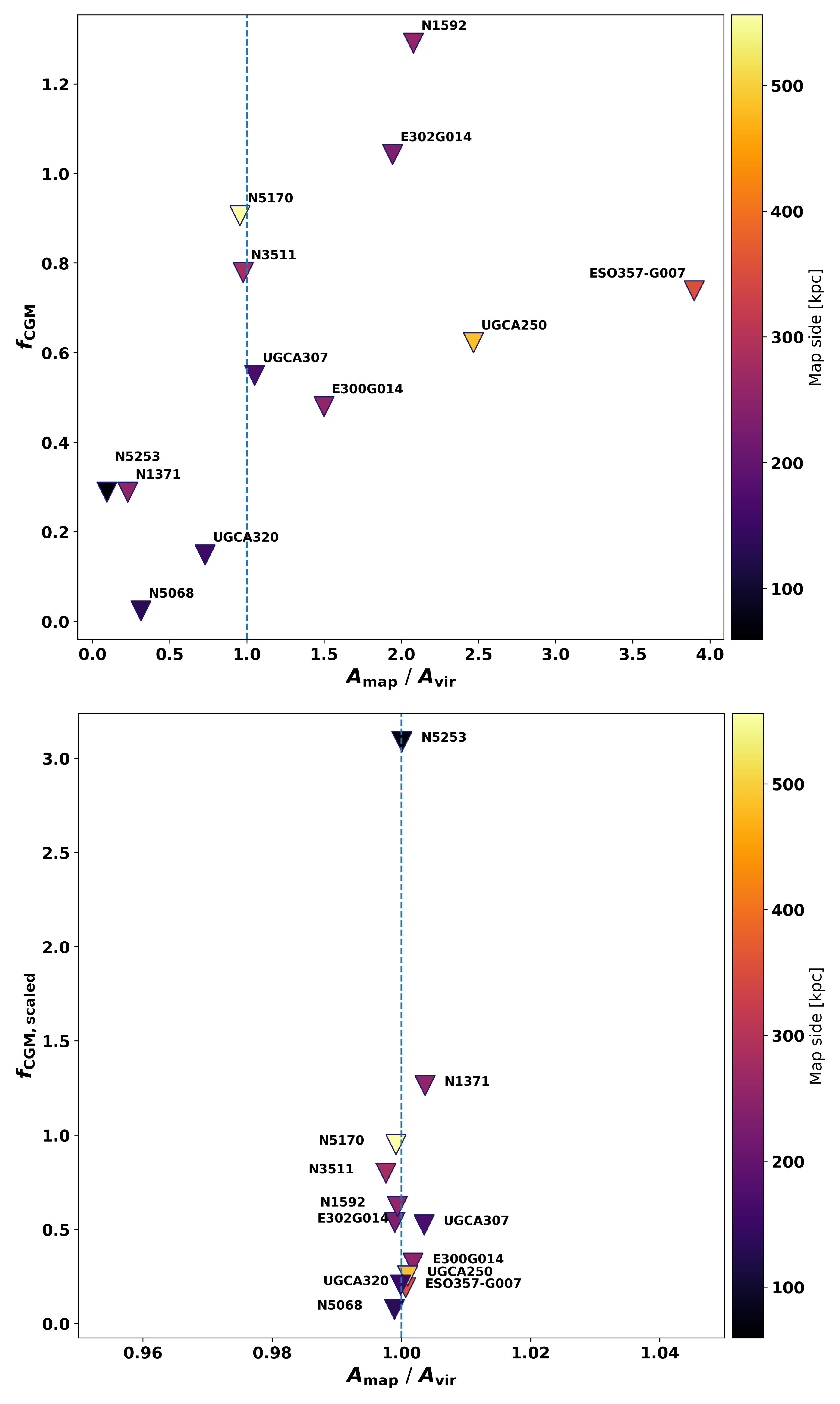}
\caption{{\it Top}. The amount of \hi detected in the CGM as a fraction of the \hi disk mass. These CGM mass measurements were taken throughout the map area and are shown relative to the target galaxy's virial area. The colorbar identifies the physical size of the map. The blue dashed line indicates a map area equal to the virial area. {\it Bottom}. The fraction of mass in the CGM in the top plot now scaled such that each map area is equal to its virial area. Triangles offset for visual ease. 
\label{figures:chapter3:fex_vs_area}}
\vspace{0.5cm}
\end{figure}

As our physical map sizes are not uniform, we would like to know how the fraction of \hi mass in the CGM would change, assuming smooth coverage of the gas, if each galaxy were mapped precisely to its virial radius, calculated from the virial mass found using the \cite{2010ApJ...710..903M} stellar-to-halo-mass relation. We show this change in Figure \ref{figures:chapter3:fex_vs_area} where our empirical measurements are shown in the top panel, and a scaled version is shown in the lower panel. We find that as an increasing fraction of the halo is probed, and in several cases multiple times the halo, the fraction of \hi gas in the CGM also increases. The largest value of $f_{CGM}$ comes from the map of NGC~1592. We made measurements over an area twice the size of its halo and found 1.3 times the amount of \hi disk mass in the surrounding area. However, when scaled the map to cover precisely one halo area, $A_{\rm map} / A_{\rm{vir}} = 1$, this fraction reduces to 0.62 times the \hi disk mass. We note that the scaled fractions, $f_{\rm CGM,scaled}$, reveal an outlier containing three times the amount of \hi mass in its CGM, NGC~5253, which we earlier identified as a candidate for accretion. We discuss this further in the following sections.

\subsection{Fraction of low \nhi}
We define a characteristic column density as low \nhi if it is below a level of $10^{19}\, \rm cm^{-2}$. This value is taken from the prediction that below $\sim 10^{19}\, \rm cm^{-2}$ the amount of hydrogen in the neutral phase is truncated, and the gas transitions from mostly neutral to mostly ionized, making detections much below this level difficult, although not impossible. This sharp drop is largely due to photoionization of the neutral hydrogen by the extragalactic radiation field and represents one way of defining the edge of an \hi disk. This was first explained by \cite{1977SvA....21..542B} and later by \cite{1993ApJ...414...41M} who suggested that the ionization fraction sharply increases at a critical column density of a few times $10^{19}\, \rm cm^{-2}$, irrespective of galaxy mass or halo parameters. An example of this sharp drop-off in \hi is the KAT-7 observations of M83, which revealed a steep decrease in \nhi at the edge of the galaxy disk \citep{2016MNRAS.462.1238H}. Recently, \cite{2017ApJ...849...51B} demonstrated that it may be possible to detect low column density \hi at significantly larger radii than the predicted truncation radius from \cite{1993ApJ...414...41M}. Specifically, they showed that the radius at which 50\% of the total amount of hydrogen is ionized is significantly larger. These results seem to be congruent with those of \cite{2018AJ....155..233I} who investigated radially averaged column density profiles of 17 galaxies in search of a break at the $\sim 10^{19}\, \rm cm^{-2}$ level and found no evidence for a sharp change anywhere above their detection limits.

We determined the diffuse neutral fraction for each galaxy using \citep{2018ApJ...865...36P}:

\begin{equation}
    f_{19} = 1 - \frac{M_{19}}{M_{\hi}} \, .
\end{equation}

\noindent In this equation, the diffuse neutral fraction, $f_{19}$, is defined by the fraction of \hi below column densities of $10^{19}\, \rm cm^{-2}$, where $M_{19}$ is the \hi mass at column density levels \nhi$\geq 10^{19}\, \rm cm^{-2}$ and $M_{\hi}$ is the total \hi mass as determined in Section \ref{subsec:analysis:cumuHI}. Values of $f_{19}$ are listed in Table \ref{tab:chapter3:tdep}. 

\begin{figure*}
\centering
\includegraphics[trim=0cm 0cm 0cm 0cm, clip,width=0.8\textwidth]{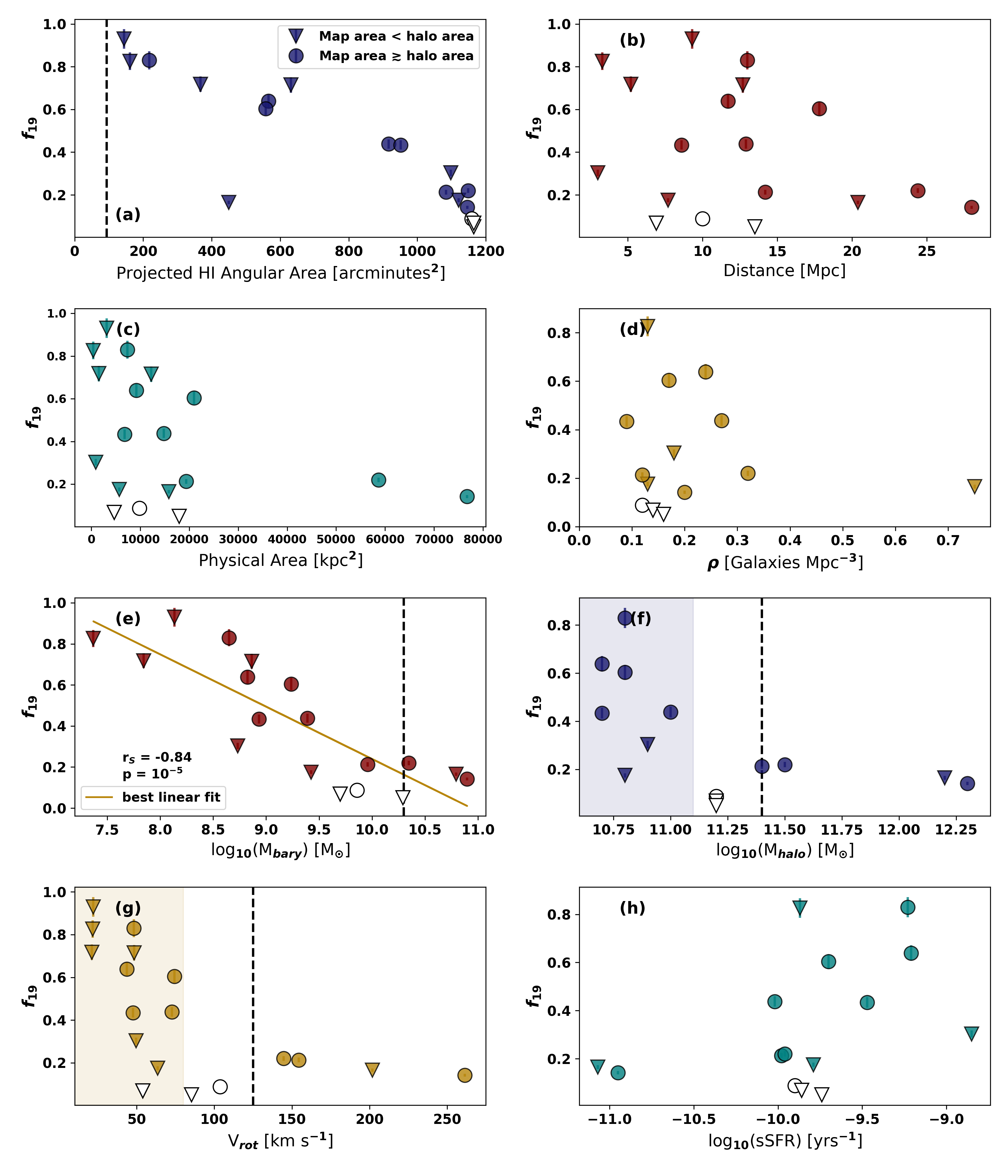}
\caption{Comparisons of the diffuse neutral fraction for each GBT-MHONGOOSE galaxy. Circles represent measurements over an area larger than or equal to the area of the halo. Triangles represent measurements over a region smaller than the halo and should be interpreted as a lower limit. The three galaxies (NGC~5068, NGC~1744, NGC~7424) whose cumulative flux profiles flatten at large radii, indicating the \hi has truncated, are shown as unfilled markers. (a) Angular areas calculated inside region where column densities reach \nhi$>1\times10^{19}\, \rm cm^{-2}$. Vertical dashed line represents the angular area of the $9.1'$ FWHM GBT beam. (b) Distance. (c) Physical areas calculated from the \nhi$=1\times10^{19}\, \rm cm^{-2}$ angular diameter. (d) Galaxy density. (e) Baryonic mass using $\rm M_{bary}=1.36\cdot M_{\hi} + M_*$. Vertical dashed line at the cold accretion threshold given in \cite{2005MNRAS.363....2K}. We show a best fit line through the data revealing, rather than a firm threshold for cold accretion, a correlation between a galaxy's baryonic mass and its diffuse gas fraction. The strength of this correlation is demonstrated with the Spearman correlation coefficient and the associated p-value. (f) Dark matter halo mass derived from the \cite{2010ApJ...710..903M} relation. Vertical dashed line at threshold also given in \cite{2005MNRAS.363....2K}. The shaded region marks a lower threshold showing, based on low \nhi measurements from this work, the regime in which cold accretion is more likely. (g) Rotation velocity using $\rm V_{rot} = W_{20}/2$. A vertical dashed line at 125 km/s shows the threshold suggested by \cite{2013ApJ...777...42K}, below which galaxies become dominated by their gas. We mark a shaded region, based on this work, demonstrating a lower threshold for a diffuse gas dominated regime below 80 km/s. (h) Specific star formation rate calculated using SFR from \cite{2019ApJS..244...24L}.
\label{fig:chapter3:f19}}
\end{figure*}

With our $f_{19}$ values in hand, we would like to compare the diffuse neutral fraction as a function of galaxy parameter, such as galaxy density, baryonic mass, and rotation velocity, and others plotted in Figure \ref{fig:chapter3:f19}. We chose these properties in order to make a comparison with theoretical predictions on the relationships between diffuse gas and cold mode accretion from the IGM. 

Our diffuse fractions range from 0.05 to 0.93 of the total \hi mass measured in each galaxy. However, as the cumulative flux profile in nearly all of our galaxies continued to rise even at larger radii, these fraction represent a lower limit. The three galaxies, NGC~1744, NGC~7424, and NGC~5068, in which the \hi truncates at some radius have a very low diffuse fraction and are unlikely to change. Eleven of our galaxies are made up of less than 50\% diffuse gas. These galaxies are generally our highest \hi mass galaxies, demonstrating a perhaps intuitive correlation between mass and diffuse fraction: higher \hi mass corresponds to a lower fraction of diffuse \hi. This correlation can be seen when we combine the stellar mass and \hi mass, or the baryonic mass, in panel (e) of Figure \ref{fig:chapter3:f19}.

In panel (a) of Figure \ref{fig:chapter3:f19}, we demonstrate the size of the galaxies in relation to the GBT $9.1'$ FWHM beam area, which is illustrated by the dashed vertical line. This helps us determine if a galaxy is large enough to be considered resolved in the GBT beam, which tells us if the \nhi detected within the beam is representative of the real \nhi or if it is spread out over the beam, diluting the physical \nhi values of the galaxy. Each galaxy's area was determined by the number of pixels within the galaxy where column density values were greater than $1\times10^{19}\, \rm cm^{-2}$. We note that the highest $f_{19}$ value, from ESO~300-16, corresponds to the map with the smallest angular area. 

At the largest distances, the GBT's beam covers the largest physical area, and could potentially be affected by beam dilution producing unphysically high values of $f_{19}$. It is clear in panel (b) that we do not see evidence of high $f_{19}$ values at large distances, or correspondingly large physical map areas. At an intermediate distance of 9.3 Mpc, beam dilution in the map of ESO~300-16 is not a concern. 

Another way to look for this bias toward unphysical low column densities is to plot $f_{19}$ against physical area, which we do in panel (c). Bias in this plot would present as large physical areas with high $f_{19}$ values, due to beam dilution. As we see nothing to indicate any such trend, we are confident that our spatial resolution is suitable for this kind of analysis. We draw attention to the two outlier galaxies in panel (c): NGC~5170 and UGCA~250. These galaxies make up the largest distances in our sample, are the first and third largest \hi disk masses in our sample, and contain less than 25\% diffuse \hi. This is consistent with the general idea that galaxies with larger potential wells are more efficient at bringing halo gas to the disk, leaving less diffuse gas in the CGM. It could also be the result of a hot halo, which is more common around massive galaxies \citep{2009MNRAS.395..160K}, and most of the gas has been ionized, resulting in a smaller neutral fraction. 

In panel (d) we look at the environments in which these galaxies live using galaxy number densities from the Nearby Galaxy Catalog \citep{1988ngc..book.....T}. This catalog identifies the density of galaxies brighter than -16 magnitudes measured in the B-band within a $\rm Mpc^{3}$ region of the target source using a 3D-grid at 0.5 Mpc spacing. We were unable to obtain galaxy number densities for four of our galaxies (ESO300-16, KK98-195, KKS2000-23, and NGC1592), reducing the size of our sample for this analysis. Regardless, the data we show in panel (d) do not indicate any relation between a galaxy's diffuse \hi fraction and its density environment.

In panel (e) we plot $f_{19}$ with baryonic mass for all but three of our galaxies. \cite{2005MNRAS.363....2K} show that cold mode accretion dominates at baryonic masses ($\rm M_{bary}=1.36\cdot M_{\hi} + M_*$) less than $\rm M_{bary}=10^{10.3}$ \msun. We calculated the baryonic masses using our derived $M_{\hi}$ disk value, which was corrected by a factor of 1.36 to account for helium mass, and stellar masses given in \cite{2019ApJS..244...24L}, which provided stellar masses for all but three of our sources. Within our sample of galaxies, we do find that galaxies with higher fractions of diffuse \hi lie entirely below this threshold, complementing the picture of diffuse gas from the IGM flowing directly into galaxies below this mass threshold. However, rather than one mass threshold below which diffuse gas may flow from the IGM, we see a relation between this galaxy mass and the diffuse \hi fraction. The trend we see in the comparison of $f_{19}$ with baryonic mass is quantified by performing tests using both the Spearman correlation coefficient and the Pearson correlation coefficient resulting in nearly identical values of $r_S \simeq r \simeq -0.84$ and the p-value, showing the probability of this correlation occurring by chance, of $p \simeq 10^{-5}$. These tests demonstrate a strong relation between the amount of diffuse \hi in a galaxy and that galaxy's baryonic mass. This observational trend complements the simulations from \citep{2009MNRAS.395..160K} which reveal a smooth trend of increasing cold gas fraction as galaxy mass decreases.

We show dark matter halo masses in panel (f), derived from the \cite{2010ApJ...710..903M} stellar-to-halo-mass relation. In this panel we see that all galaxies with halo masses below $\sim 10^{11.1}$ \msun{} have a spread in diffuse \hi fractions, while above this threshold the diffuse fraction remains $\lesssim 0.2$. This data indicates that there may be a dark matter halo mass threshold around $10^{11.1}$ \msun, above which a galaxy contains less than 25\% diffuse \hi. We have overlaid a vertical dashed line at the predicted threshold where cold accretion should dominate below $\rm M_{halo}=10^{11.4}$ \msun{} given in \cite{2005MNRAS.363....2K}. Below this halo mass threshold, we should find larger diffuse fractions associated with those galaxies. The shaded region in panel (f) pushes back this threshold to below $\sim 10^{11.1}$ \msun{}.

Rotation velocities are plotted against $f_{19}$ in panel (g) with a distinction being made at the $125\, \rm km\, s^{-1}$ mark. \cite{2013ApJ...777...42K} suggests that this is the threshold below which galaxies are gas-dominated. In this regime, galaxies are refueled and have larger gas fractions than those above this threshold. We plot our alternative measurement using $f_{19}$, which would estimate a ``diffuse" gas richness. Once again, while we have a spread of diffuse fraction values below $v_{rot} = 125\, \rm km\, s^{-1}$, galaxies with higher $f_{19}$ values all lie below this threshold supporting the picture of diffuse gas fueling galaxies. We further demonstrate that the data suggests an even lower threshold where galaxies potentially become dominated by their diffuse gas. This is marked in panel (g) by the shaded below a rotation velocity of 80 km/s. 

The last panel (h) in Figure \ref{fig:chapter3:f19} is of specific star formation rate (sSFR) versus $f_{19}$. We derive sSFR, the star formation rate per unit stellar mass, using the same stellar masses mentioned above with star formation rates derived from UV+IR provided by \cite{2019ApJS..244...24L}. As star formation and galaxy mass is directly related to cold mode accretion, we want to look for trends in sSFR as it relates to both. We note in this panel that the highest diffuse fractions tend to fall at the higher end of our sSFR range, which could be contributed to by either inflows into or outflows from the galaxy (e.g. \cite{2010ApJ...719.1503R, 2012ApJ...760..127M, 2012ApJ...753...16K}). 

\begin{deluxetable}{lcccr}
\tabletypesize{\normalsize}
\tablecaption{\hi Depletion Timescales}
\setlength{\tabcolsep}{5pt} 
\renewcommand{\arraystretch}{1.25} 
\tablewidth{0pt}

\tablehead{
\colhead{Source} &
\colhead{SFR} &
\colhead{log $M_{disk}^{\hi}$} &
\colhead{$\tau_{disk}^{gas}$} &
\colhead{$f_{19}$} \\
\colhead{} &
\colhead{[\msun $\rm yr^{-1}$]} &
\colhead{[\msun]} &
\colhead{[Gyr]} 
}

\decimalcolnumbers
\startdata
E300G014 & 0.063 & 9.12 & 28.41 & 0.44 \\
E300G016 & ... & 8.0 & ... & 0.93 \\
E302G014 & 0.038 & 8.65 & 15.98 & 0.64 \\
ESO357-G007 & 0.041 & 9.05 & 37.46 & 0.60 \\
KK98-195 & ... & 7.71 & ... & 0.72 \\
KKS2000-23 & ... & 8.73 & ... & 0.72 \\
N1371 & 0.427 & 9.95 & 28.41 & 0.17 \\
N1592 & 0.079 & 8.36 & 3.92 & 0.83 \\
N1744 & 0.195 & 9.62 & 29.08 & 0.09 \\
N3511 & 0.447 & 9.55 & 10.8 & 0.21 \\
N5068 & 0.288 & 9.33 & 10.08 & 0.07 \\
N5170 & 0.589 & 10.29 & 45.03 & 0.14 \\
N5253 & 0.525 & 8.09 & 0.32 & 0.30 \\
N7424 & 0.372 & 10.11 & 47.16 & 0.05 \\
UGCA015 & 0.001 & 7.12 & 24.18 & 0.83 \\
UGCA250 & 0.603 & 10.09 & 27.77 & 0.22 \\
UGCA307 & 0.031 & 8.75 & 24.75 & 0.43 \\
UGCA320 & 0.021 & 9.27 & 121.21 & 0.18 \\
\enddata
\tablecomments{Star formation rates obtained from \cite{2019ApJS..244...24L}. Column 3 lists the \hi mass of the disk. Column 4 lists the gas depletion timescales for the disk gas, including a correction for Helium mass. Column 5 lists the diffuse neutral fraction of gas below \nhi $< 10^{19}\, \rm cm^{-2}$. }
\label{tab:chapter3:tdep}
\end{deluxetable}

\subsection{Depletion timescales}
One method of determining whether nearby galaxies contain enough gas to continue to fuel star formation is to calculate the depletion timescale of the galaxy, given the gas mass and the SFR. This timescale determines how long it will take the galaxy to use up all of its gas if all of this gas is converted into stars. Our depletion timescales are calculated with:
\begin{equation}
    \tau_{disk}^{gas} = \frac{M_{disk}^{gas}}{SFR} \, ,
\end{equation}
\noindent where $\tau_{disk}^{gas}$ is the depletion timescale given the gas mass from the disk, and $M_{disk}^{gas}$ is the \hi mass from the disk, measured from the total integrated \hi profiles, corrected for helium with a factor of 1.36. These timescales, listed in Table \ref{tab:chapter3:tdep}, have a large range of values which fall between 0.32 Gyr and 121 Gyr, with a median depletion time of 27.8 Gyr. The depletion timescales we find are consistent with the \hi depletion timescales found by both \cite{2010AJ....140.1194B} and \cite{2014MNRAS.445.1392R}. These results tell us that these galaxies currently have enough gas available to them to sustain star formation for well over a Hubble time, with the exception of NGC~1592, NGC~3511, NGC~5068, and NGC~5253 whose values are less than 13 Gyr. In Section \ref{subsec:bin2} we noted NGC~5253 as a potential source for accretion. NGC~5253 has been shown to have extremely efficient star formation and streams of metal-enriched dense gas flowing into the starbursting dwarf galaxy \citep{2015Natur.519..331T}. If NGC~5253 is undergoing ongoing accretion, this along with its star formation efficiency could explain the lack of an established reservoir of diffuse \hi, as well as its shorter depletion timescale. 

\section{Summary}
\label{sec:chapter3:summary}
We analyzed 21 cm data from GBT observations of 18 MHONGOOSE galaxies to search for diffuse, low column density \hi in the CGM of these galaxies. We reached a mean $1\sigma$ column density sensitivity of \nhi = 1.3 $\times 10^{17}\, \rm cm^{-2}$ per channel, and a $3\sigma$ column density of \nhi $= 1.3 \times 10^{18}\, \rm cm^{-2}$ over a linewidth of 20 $\rm km \, s^{-1}$.

\begin{enumerate}
    \item By comparing the total \hi mass in our maps to the \hi measured in just the disk of each galaxy, we determined that 16 out of 18 galaxies contained additional \hi mass outside of their disks by an amount of 0.02-3 times as much. This extraplanar or CGM \hi was not found in NGC~1744 or NGC~7424. 
    \item We measured the amount of diffuse \hi within each map, defined as \hi gas with column densities below \nhi$= 10^{19}\, \rm cm^{-2}$. We compared the amount of diffuse \hi with the total \hi mass in the map. We found that the galaxies in our sample were made up of between 5 - 93\% diffuse \hi. If we bin our diffuse fraction values, $f_{19}$, by quartiles we find that more galaxies fall into the lowest quartile than any other bin. 
    \item We took measurements of azimuthally averaged \hi column densities around each galaxy extending out to the edges of the uniform noise regions in our maps. In eleven of our galaxies, these averaged \nhi values remained nearly constant outside the disk region, showing no excess of \hi in the averaged annuli throughout the CGM. However, we identified seven galaxies where this column density profile showed some $>5\sigma$ increase in averaged \nhi outside of the galaxy disk. These galaxies are KKS2000-23, NGC~1744, NGC~3511, NGC~5170, NGC~7424, UGCA250, and UGCA320. There are a number of reasons to find a high averaged \nhi value outside of the disk, which include detection of a companion (e.g. UGCA~320), or filamentary structure possibly associated with accretion (e.g. NGC~5170). 
    \item We see a rising cumulative flux level at increasing distances in the maps of 15 of our 18 galaxies. This is consistent with our measurement of the \hi mass outside the disk, where we find the same three galaxies (NGC~1744, NGC~7424, and NGC~5068) do not contain any detectable amount of \hi mass beyond their disks. This excess mass is seen in the cumulative flux plots where the flux does not flatten at large distances from the galaxy as might be expected if the \hi emission drops to zero.
    \item We compared the fraction of diffuse \hi in each map with various environmental properties of our galaxies in Figure \ref{fig:chapter3:f19}. We look at particular properties such as baryonic mass, dark matter halo mass, and rotation velocities which have been associated, either empirically or theoretically, with accretion of gas onto galaxies. Most notably, we find a strong correlation between the fraction of diffuse gas and the galaxy's baryonic mass. Using both the Spearman and the Pearson correlation coefficients, we find that these two properties are strongly correlated, and have a probability of a chance correlation of nearly zero. 
    \item Simulations such as seen in \cite{2005MNRAS.363....2K} determine a dark matter halo mass threshold below which cold-mode accretion is the dominate form of accretion of gas from the IGM onto galaxies. Our results indicate a lower threshold in the dark matter halo mass of a galaxy below which we find all of our high diffuse \hi fractions. Our dark matter halo mass threshold lies around $\sim 10^{11.1}$ \msun. This tells us that we can expect a galaxy with a dark matter halo mass above this threshold to contain no more than about 25\% diffuse gas. 
    \item We demonstrate that galaxies with high diffuse \hi fractions all have rotation velocities below about 80 km/s. We suggest this as a regime in which galaxies become dominated by their diffuse gas.
    \item We use the \hi mass of the disks of each galaxy, along with their star formation rates, to find the depletion timescales of each galaxy. These depletion times have a large spread, but only four galaxies have depletion times below a Hubble time. Most galaxies in our sample currently have enough \hi to continue to fuel star formation for well over a Hubble, where the median time is 27 Gyr. 
\end{enumerate}

The high resolution MHONGOOSE survey with MeerKAT will provide additional insight into the origins of low column density gas in the CGM of these galaxies. Galaxies with diffuse fractions supporting cold mode accretion models in our sample will be better understood with higher resolution data.  With this single-dish counterpart data to the MeerKAT interferometer data, we will have a complete census of the \hi in each of these galaxies at all angular scales. 

\acknowledgments
A.E.S. is supported by an NSF Astronomy and Astrophysics Postdoctoral Fellowship under award AST-1903834. D.J.P. and A.E.S. acknowledge partial support by NSF CAREER grant AST-1149491. This project has received funding from the European Research Council (ERC) under the European Union’s Horizon 2020 research and innovation programme grant agreement no. 882793, project name MeerGas. This research made use of the NASA/IPAC Extragalactic Database (NED). The Robert C. Byrd Green Bank Telescope is operated by the Green Bank Observatory. The Green Bank Observatory is a facility of the National Science Foundation operated under cooperative agreement by Associated Universities, Inc.

\bibliography{mybib.bib}
\bibliographystyle{aasjournal}

\appendix

Below we present the remainder of the galaxies in our GBT-MHONGOOSE sample, as we did in Figure \ref{fig:chapter3:NGC7424_4x4}. {\it Top left}: Total integrated \hi profile measured used to measure the total integrated \hi mass. {\it Top right}: Cumulative \hi mass. The total \hi mass in the moment 0 map is plotted by the fraction of gas in column density bins, and compared to the GBT beam model, scaled to the peak column density. {\it Lower left}: Azimuthally averaged \nhi. Column density averaged over annuli extending radially from the center of the galaxy and compared to the GBT beam model. The black dashed line characterizes the $1\sigma$ noise in each annulus, and the blue dot-dash line represents the $5\sigma$ noise in each annulus. {\it Lower right}: Cumulative flux. Flux in each of those annuli are summed to obtain a measure of the total flux out to the edge of each map.

\begin{figure*}
\centering
\includegraphics[trim=0cm 0cm 0cm 0cm, clip,width=\textwidth]{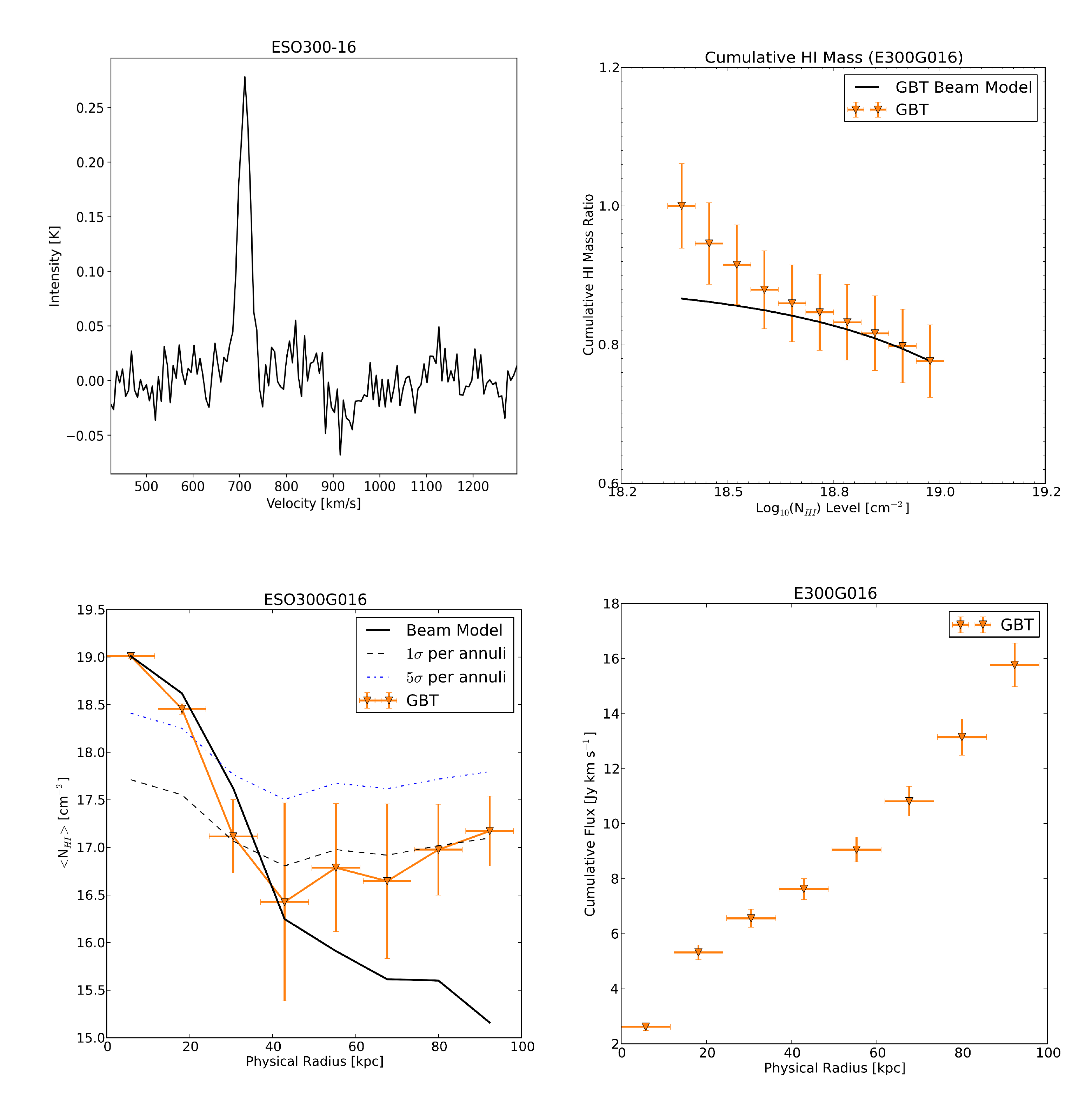}
\caption{Same as Figure \ref{fig:chapter3:NGC7424_4x4} for ESO~300-16.
\label{fig:chapter3:ESO300-G016}}
\clearpage
\end{figure*}

\begin{figure*}
\centering
\includegraphics[trim=0cm 0cm 0cm 0cm, clip,width=\textwidth]{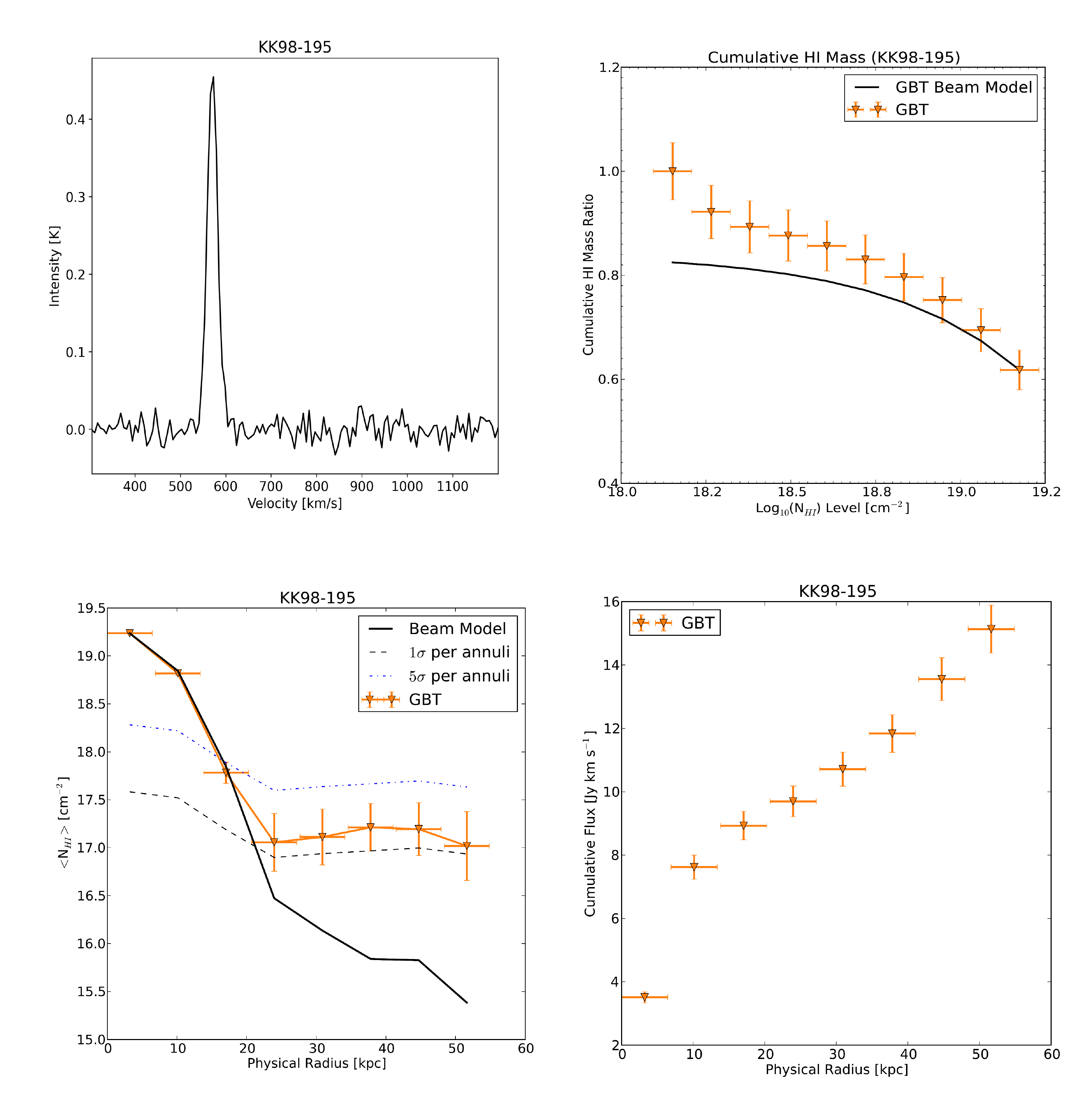}
\caption{Same as Figure \ref{fig:chapter3:NGC7424_4x4} for KK98-195. 
\label{fig:chapter3:KK98-195}}
\clearpage
\end{figure*}

\begin{figure*}
\centering
\includegraphics[trim=0cm 0cm 0cm 0cm, clip,width=\textwidth]{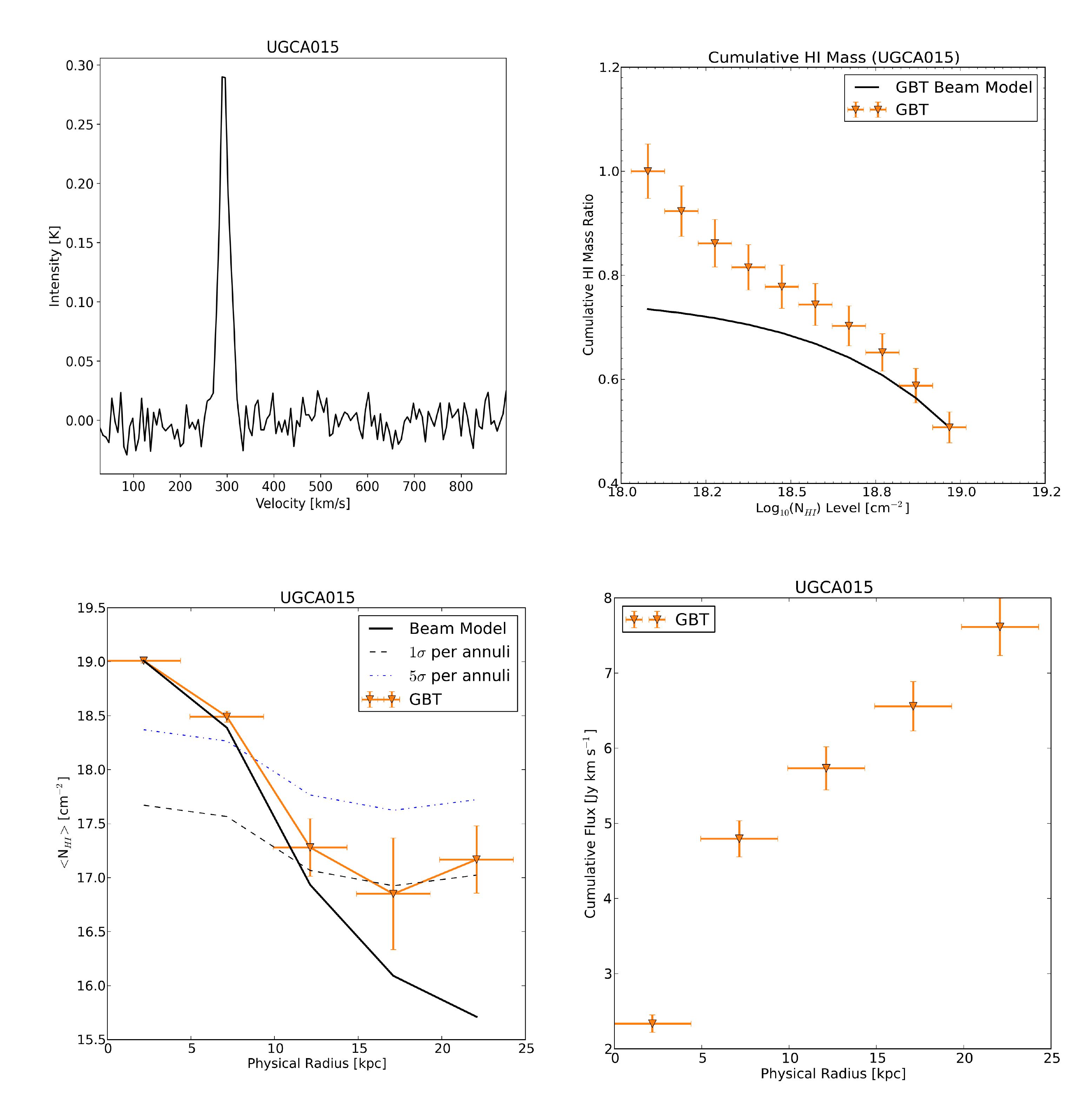}
\caption{Same as Figure \ref{fig:chapter3:NGC7424_4x4} for UGCA015. 
\label{fig:chapter3:UGCA015}}
\clearpage
\end{figure*}

\begin{figure*}
\centering
\includegraphics[trim=0cm 0cm 0cm 0cm, clip,width=\textwidth]{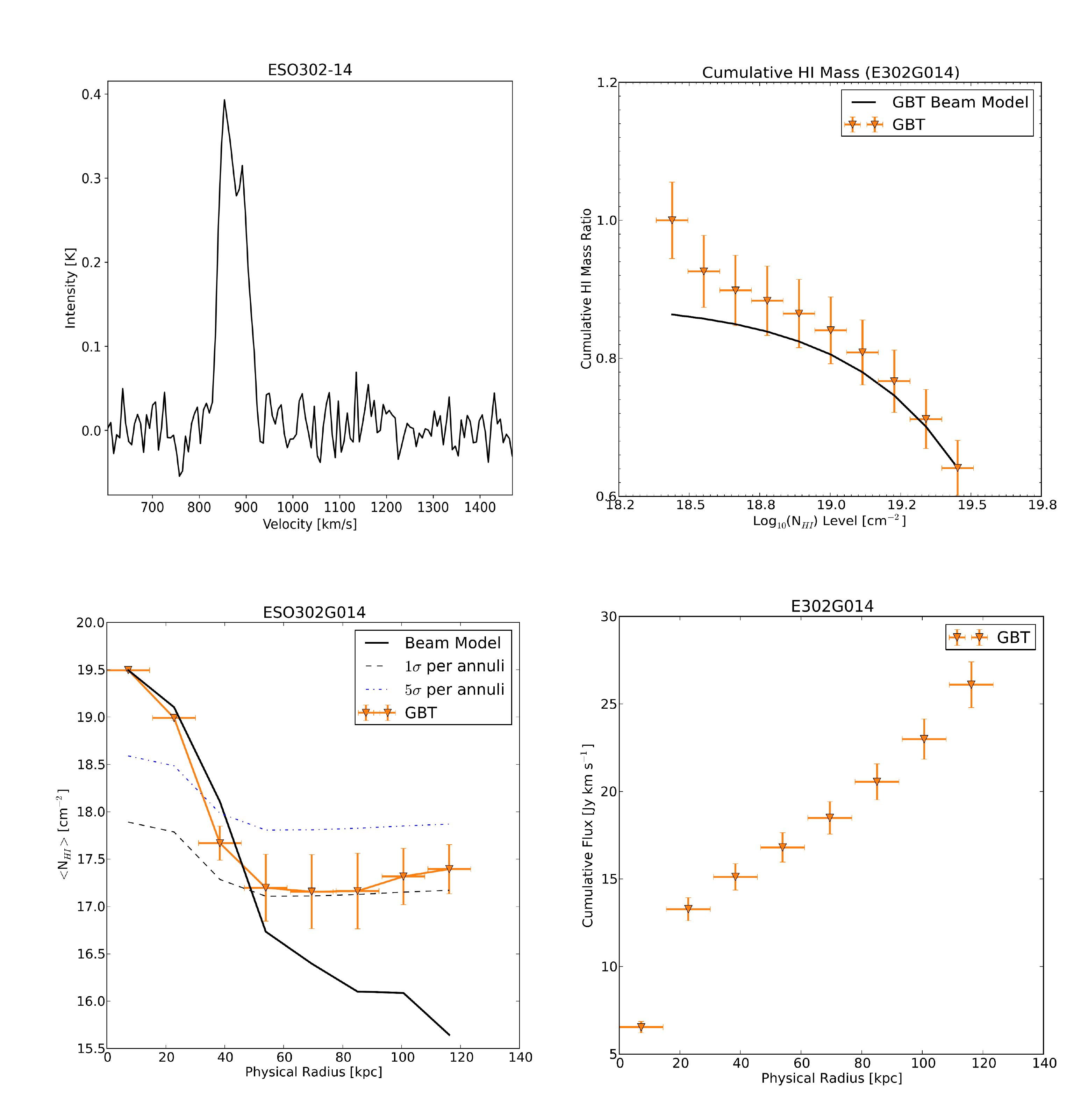}
\caption{Same as Figure \ref{fig:chapter3:NGC7424_4x4} for ESO~302-14. 
\label{fig:chapter3:ESO302-G014}}
\clearpage
\end{figure*}

\begin{figure*}
\centering
\includegraphics[trim=0cm 0cm 0cm 0cm, clip,width=\textwidth]{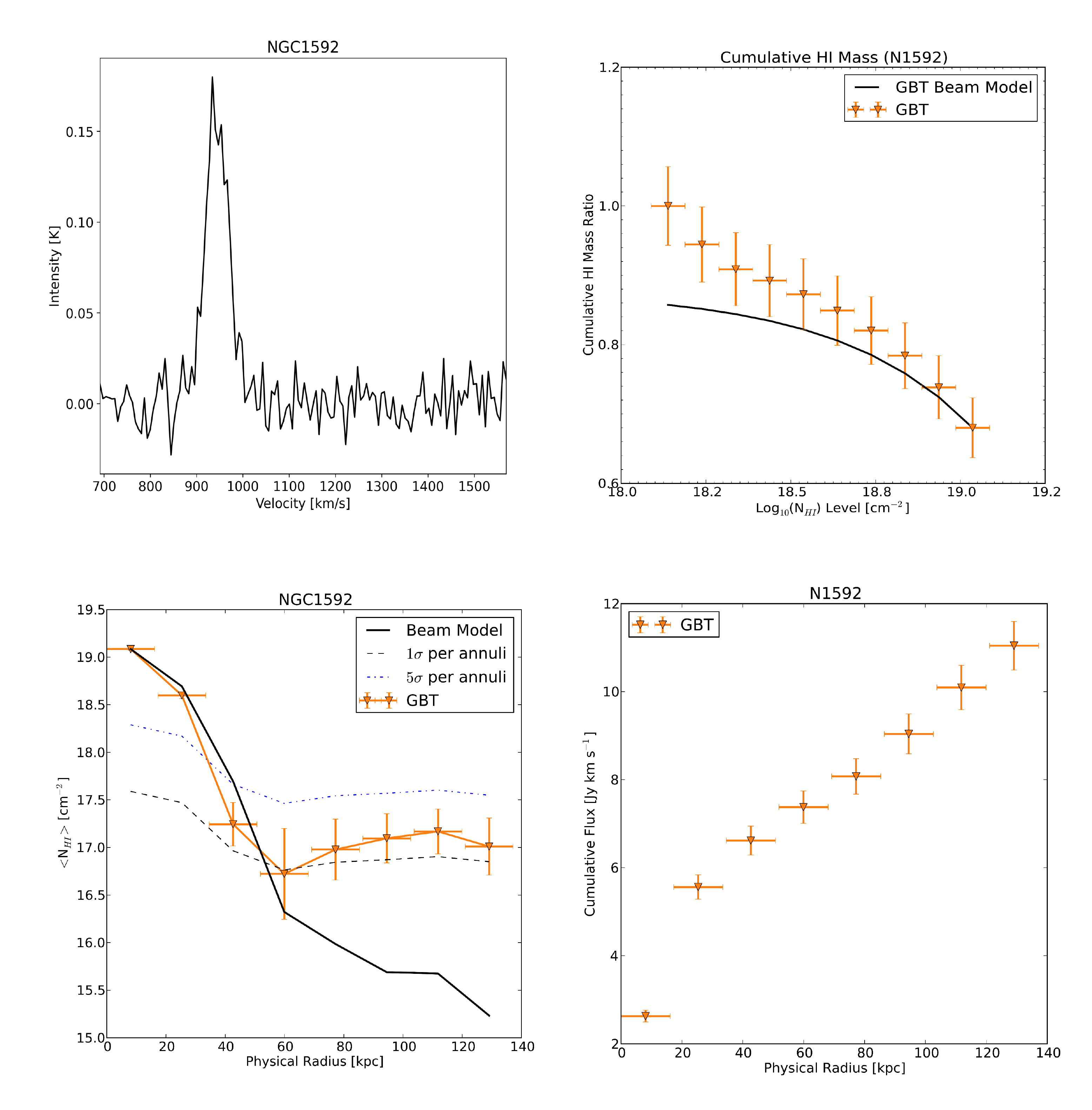}
\caption{Same as Figure \ref{fig:chapter3:NGC7424_4x4} for NGC~1592. 
\label{fig:chapter3:NGC1592}}
\clearpage
\end{figure*}

\begin{figure*}
\centering
\includegraphics[trim=0cm 0cm 0cm 0cm, clip,width=\textwidth]{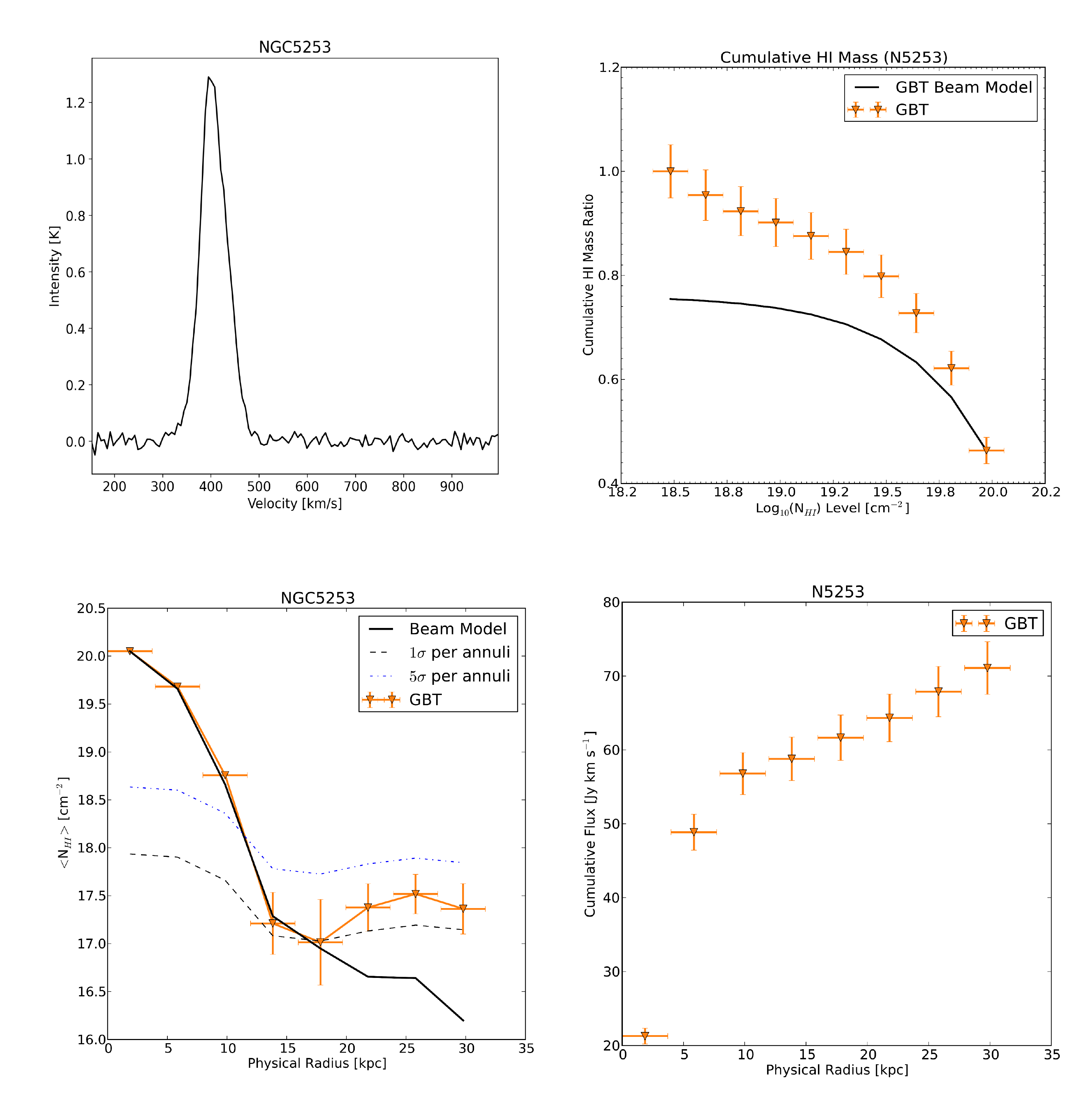}
\caption{Same as Figure \ref{fig:chapter3:NGC7424_4x4} for NGC~5253.
\label{fig:chapter3:NGC5253}}
\clearpage
\end{figure*}

\begin{figure*}
\centering
\includegraphics[trim=0cm 0cm 0cm 0cm, clip,width=\textwidth]{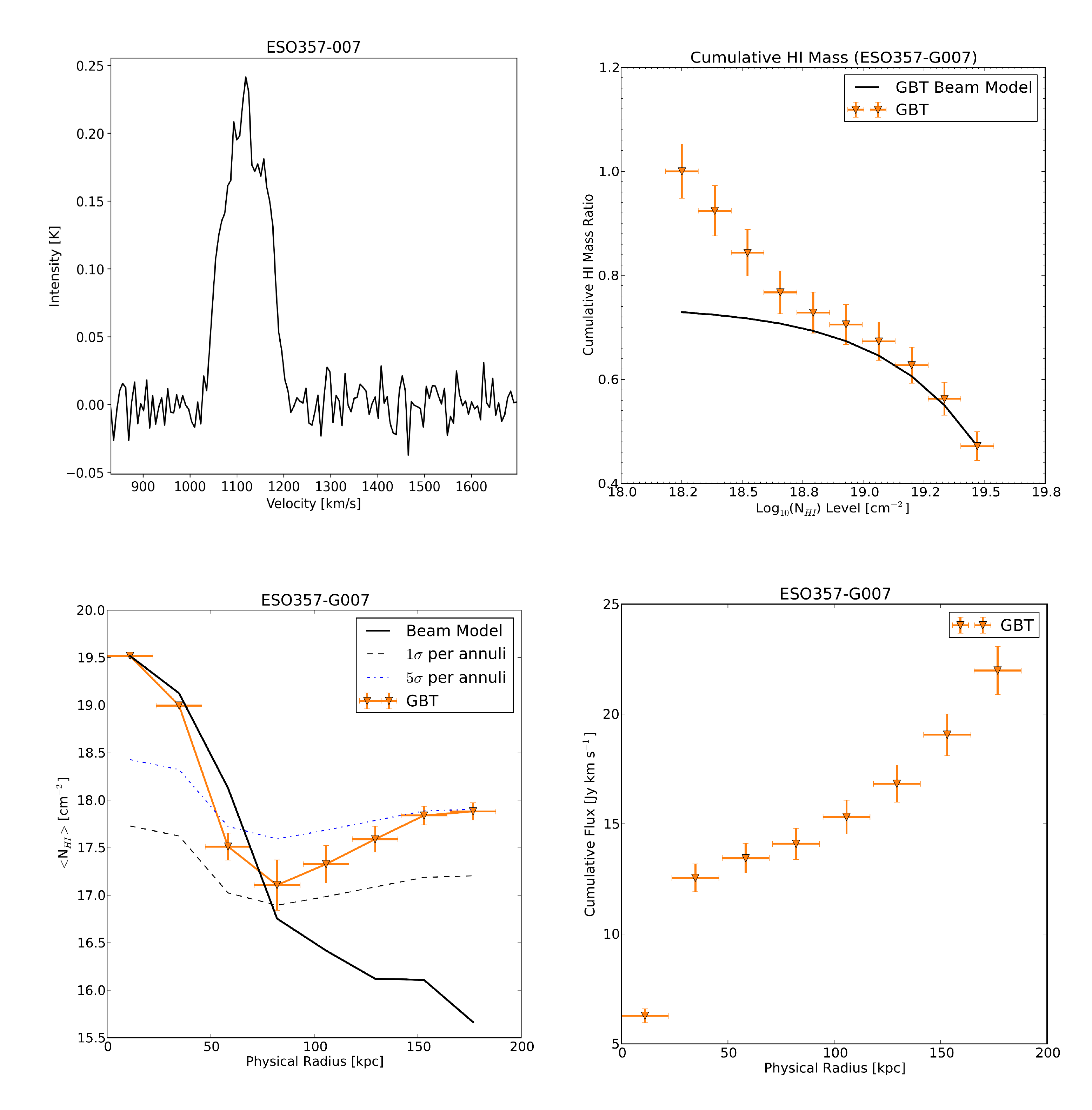}
\caption{Same as Figure \ref{fig:chapter3:NGC7424_4x4} for ESO~357-007. 
\label{fig:chapter3:ESO357-G007}}
\clearpage
\end{figure*}

\begin{figure*}
\centering
\includegraphics[trim=0cm 0cm 0cm 0cm, clip,width=\textwidth]{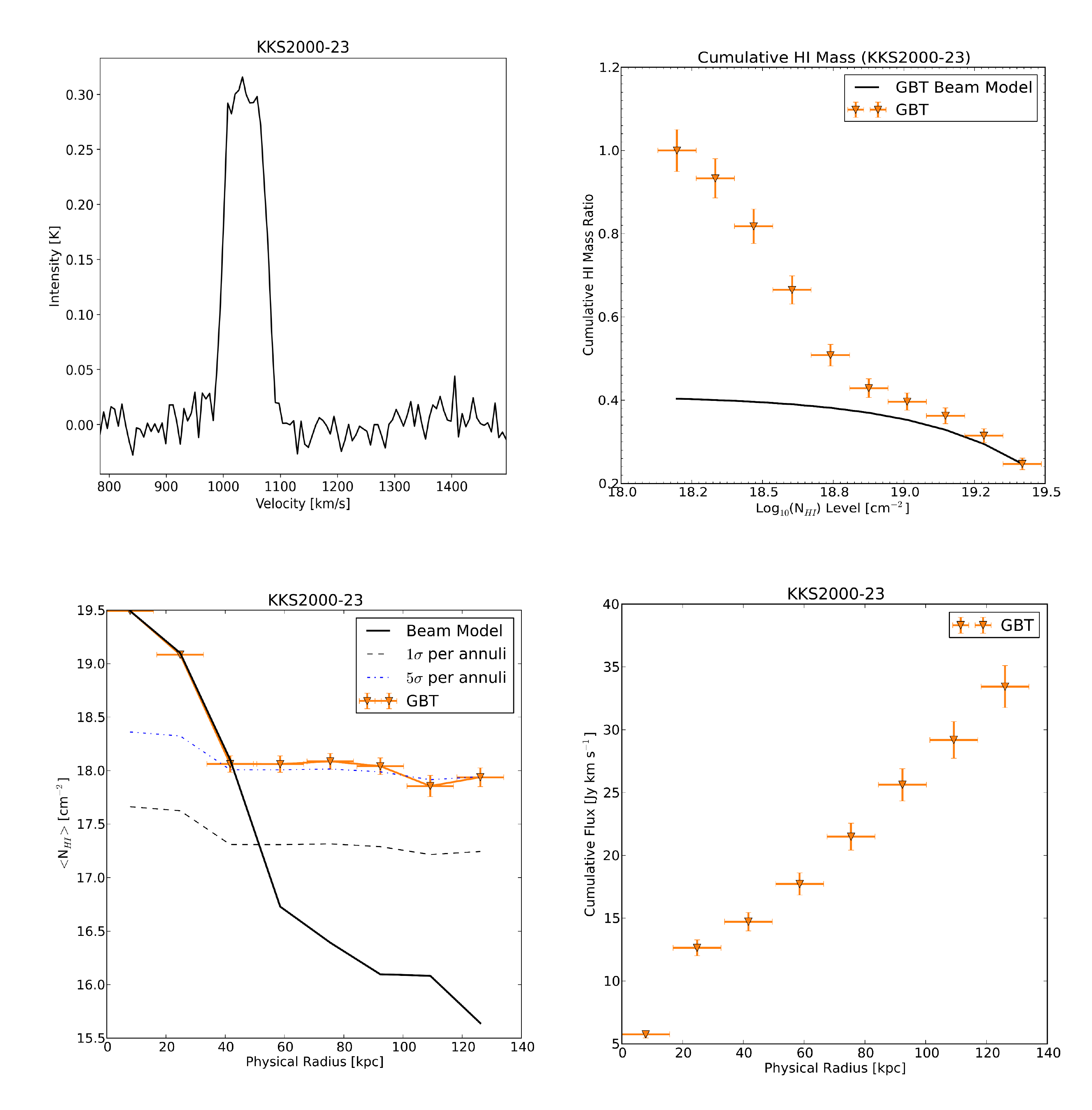}
\caption{Same as Figure \ref{fig:chapter3:NGC7424_4x4} for KKS2000-23. 
\label{fig:chapter3:KKS2000-23}}
\clearpage
\end{figure*}

\begin{figure*}
\centering
\includegraphics[trim=0cm 0cm 0cm 0cm, clip,width=\textwidth]{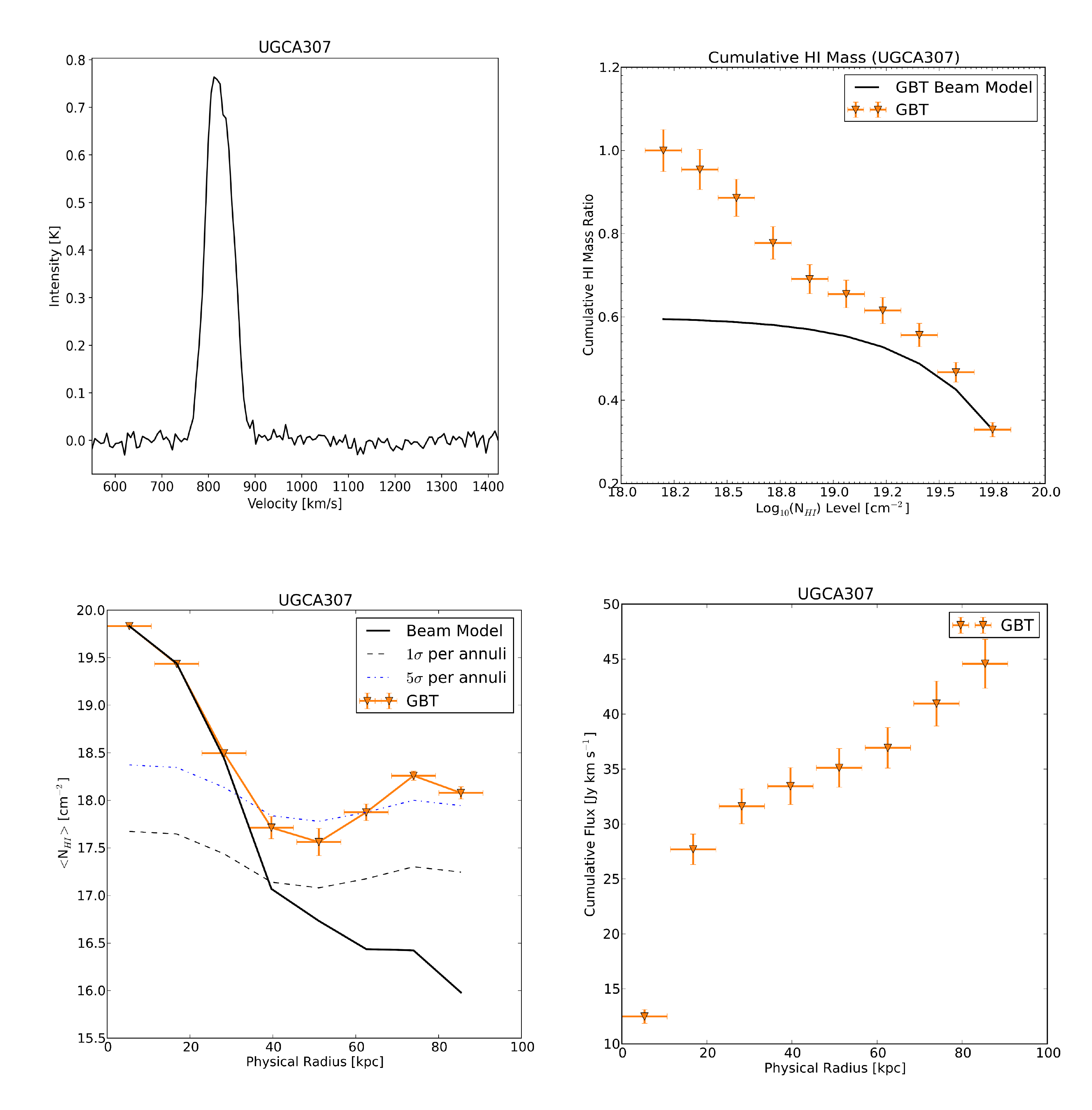}
\caption{Same as Figure \ref{fig:chapter3:NGC7424_4x4} for UGCA307. 
\label{fig:chapter3:UGCA307}}
\clearpage
\end{figure*}

\begin{figure*}
\centering
\includegraphics[trim=0cm 0cm 0cm 0cm, clip,width=\textwidth]{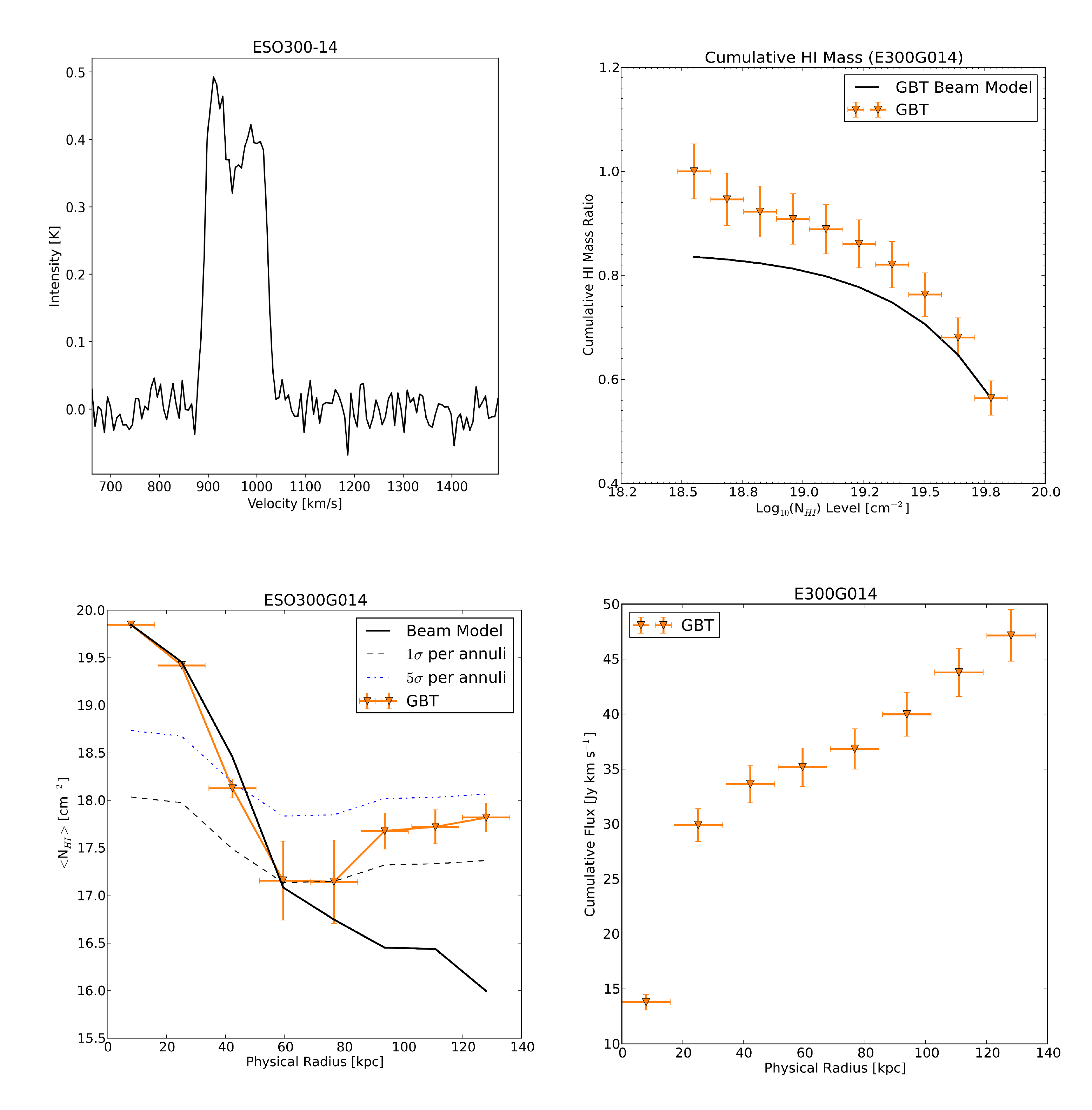}
\caption{Same as Figure \ref{fig:chapter3:NGC7424_4x4} for ESO300-14. 
\label{fig:chapter3:ESO300-G014}}
\clearpage
\end{figure*}

\begin{figure*}
\centering
\includegraphics[trim=0cm 0cm 0cm 0cm, clip,width=\textwidth]{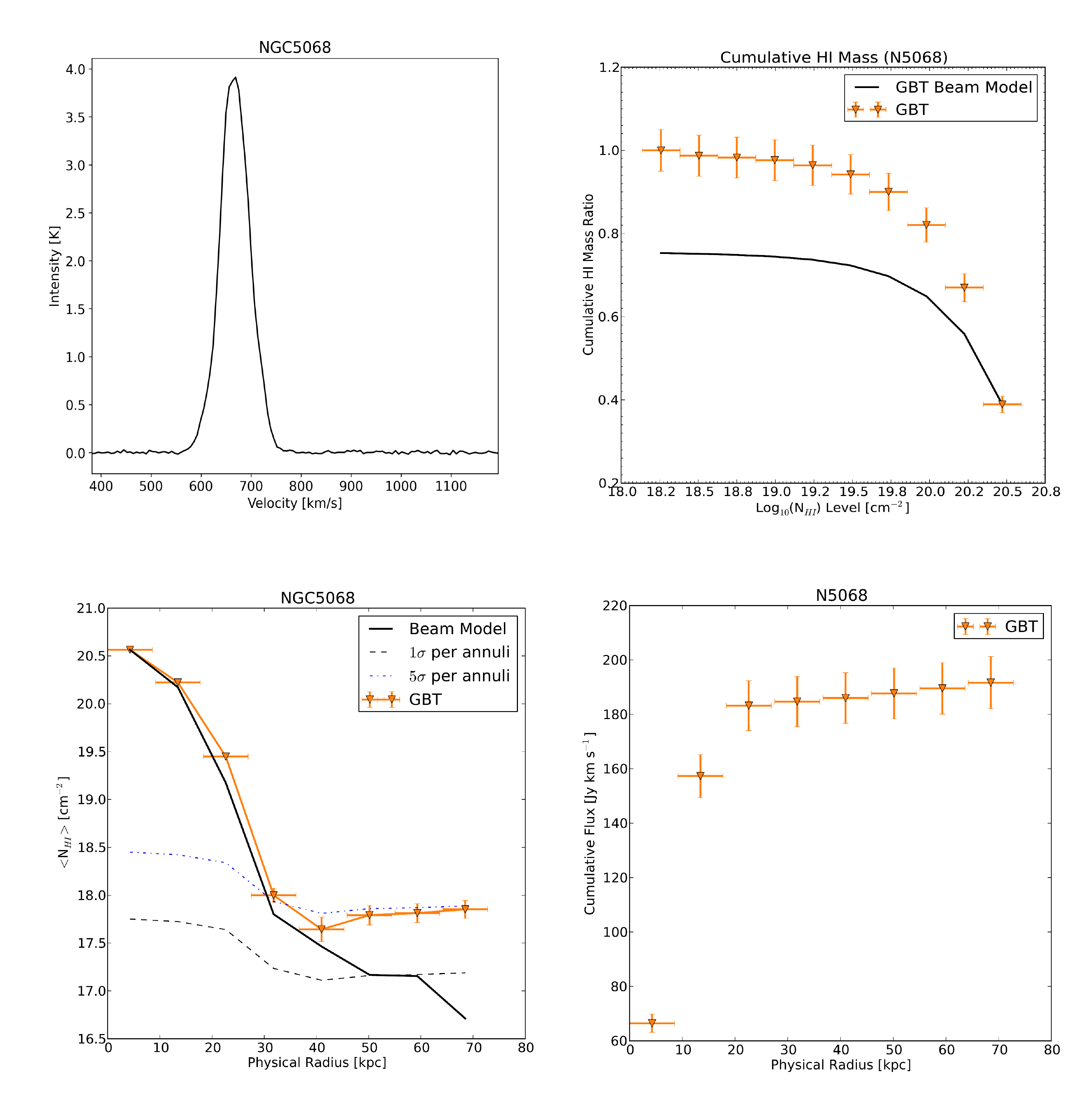}
\caption{Same as Figure \ref{fig:chapter3:NGC7424_4x4} for NGC~5068. 
\label{fig:chapter3:NGC5068}}
\clearpage
\end{figure*}

\begin{figure*}
\centering
\includegraphics[trim=0cm 0cm 0cm 0cm, clip,width=\textwidth]{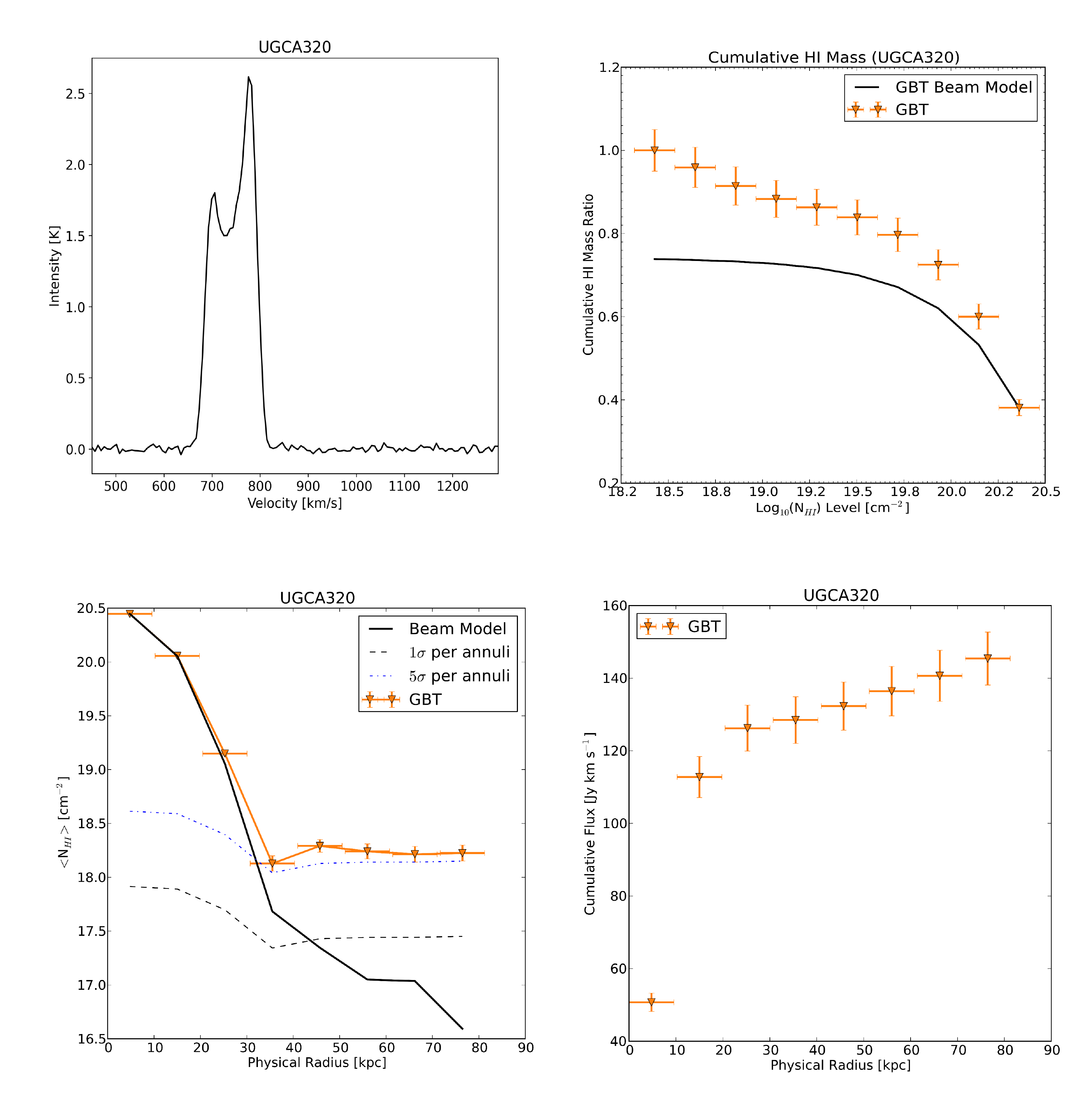}
\caption{Same as Figure \ref{fig:chapter3:NGC7424_4x4} UGCA320. 
\label{fig:chapter3:UGCA320}}
\clearpage
\end{figure*}

\begin{figure*}
\centering
\includegraphics[trim=0cm 0cm 0cm 0cm, clip,width=\textwidth]{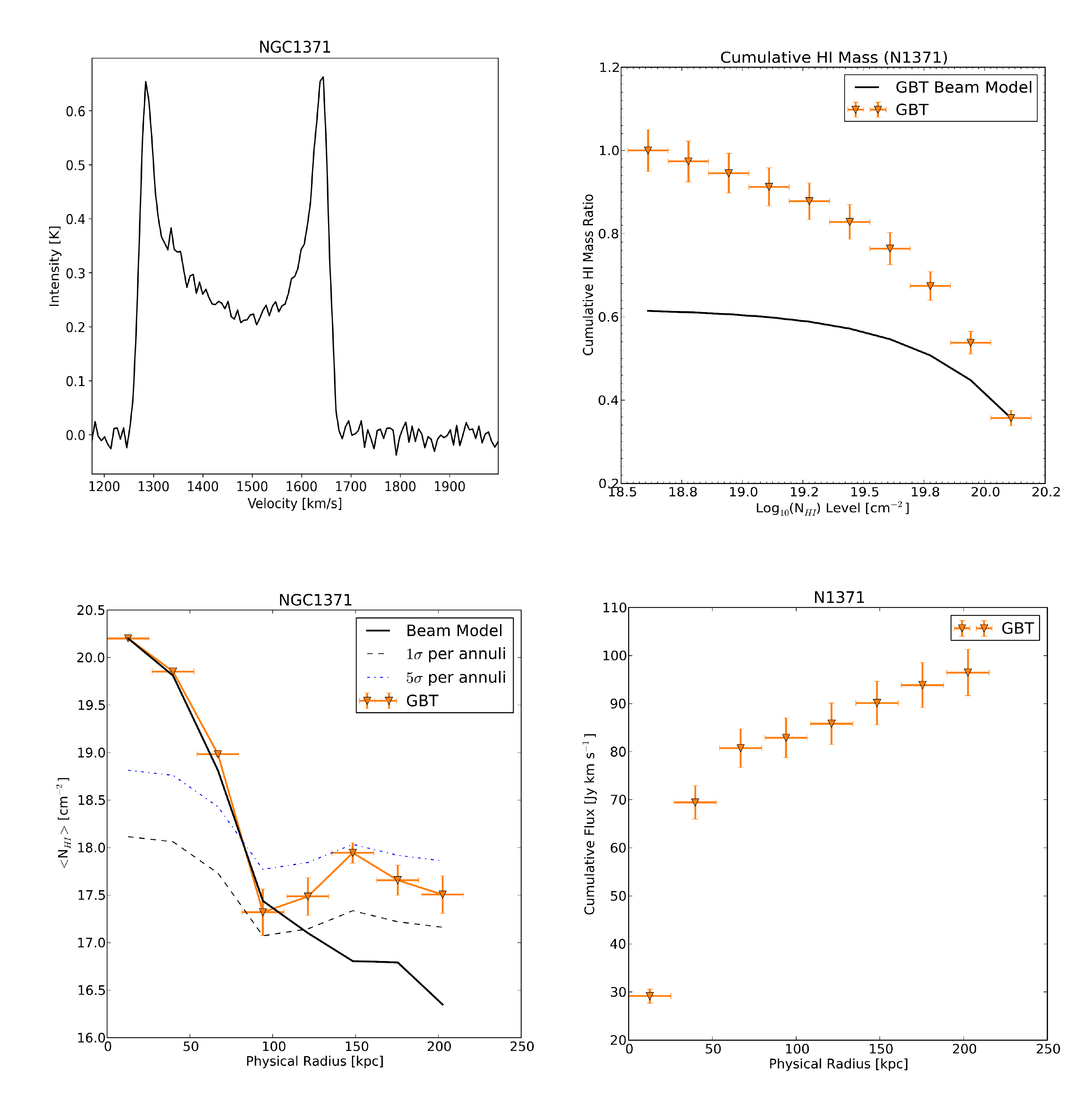}
\caption{Same as Figure \ref{fig:chapter3:NGC7424_4x4} for NGC~1371. 
\label{fig:chapter3:NGC1371}}
\clearpage
\end{figure*}

\begin{figure*}
\centering
\includegraphics[trim=0cm 0cm 0cm 0cm, clip,width=\textwidth]{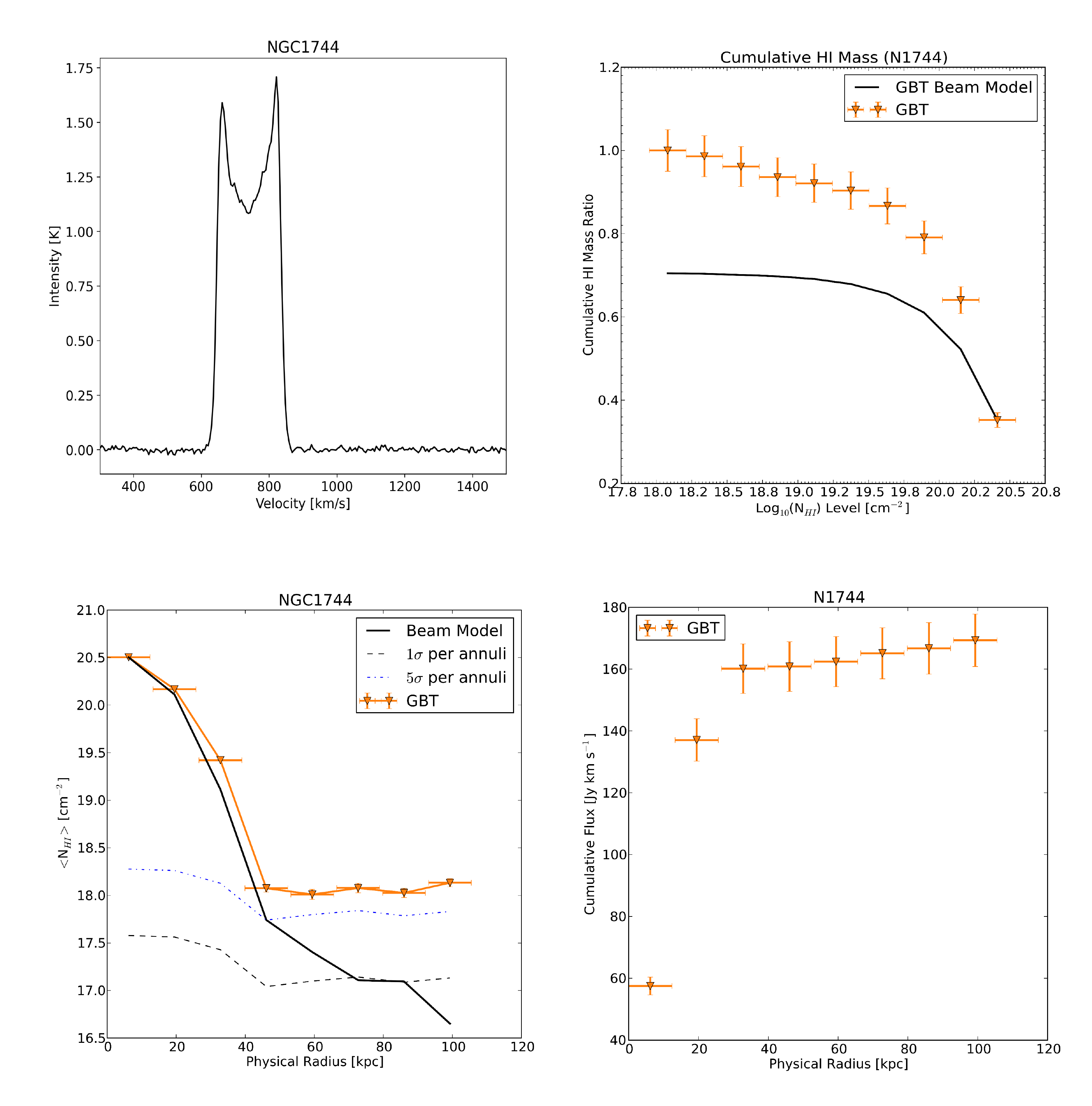}
\caption{Same as Figure \ref{fig:chapter3:NGC7424_4x4} for NGC~1744. 
\label{fig:chapter3:NGC1744}}
\clearpage
\end{figure*}

\begin{figure*}
\centering
\includegraphics[trim=0cm 0cm 0cm 0cm, clip,width=\textwidth]{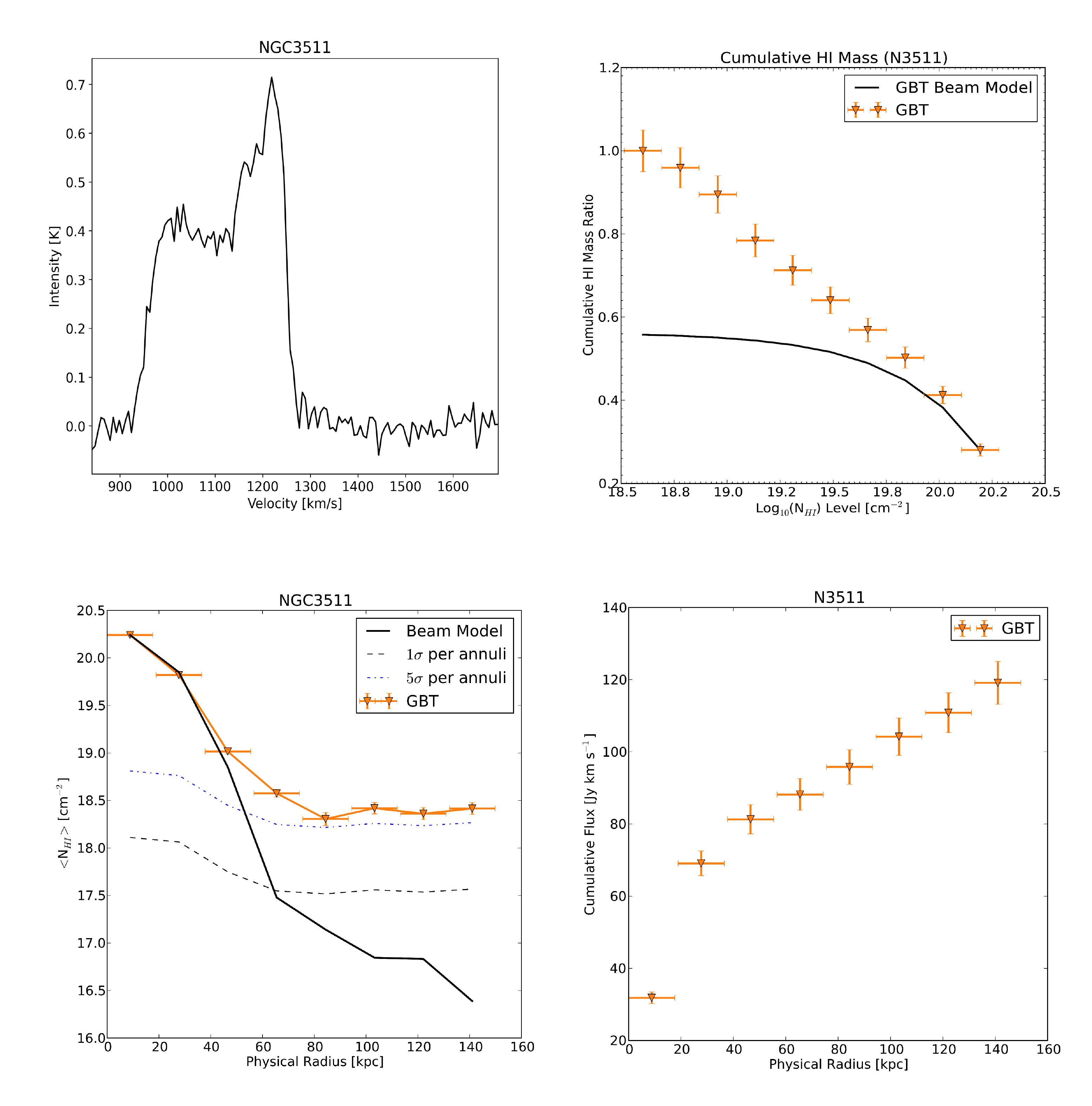}
\caption{Same as Figure \ref{fig:chapter3:NGC7424_4x4} for NGC~3511. 
\label{fig:chapter3:NGC3511}}
\clearpage
\end{figure*}

\begin{figure*}
\centering
\includegraphics[trim=0cm 0cm 0cm 0cm, clip,width=\textwidth]{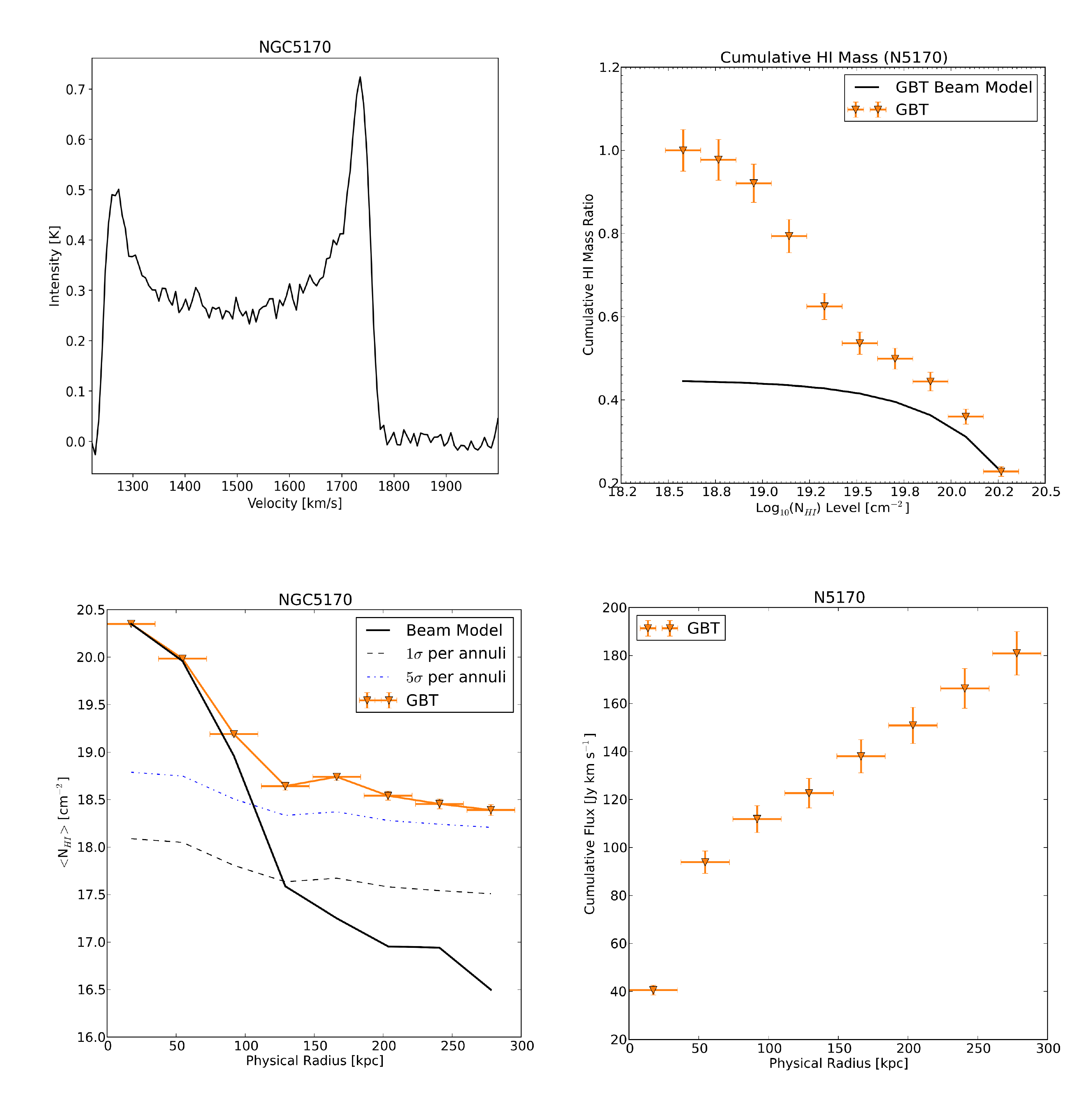}
\caption{Same as Figure \ref{fig:chapter3:NGC7424_4x4} for NGC~5170. 
\label{fig:chapter3:NGC5170}}
\clearpage
\end{figure*}


\begin{figure*}
\centering
\includegraphics[trim=0cm 0cm 0cm 0cm, clip,width=\textwidth]{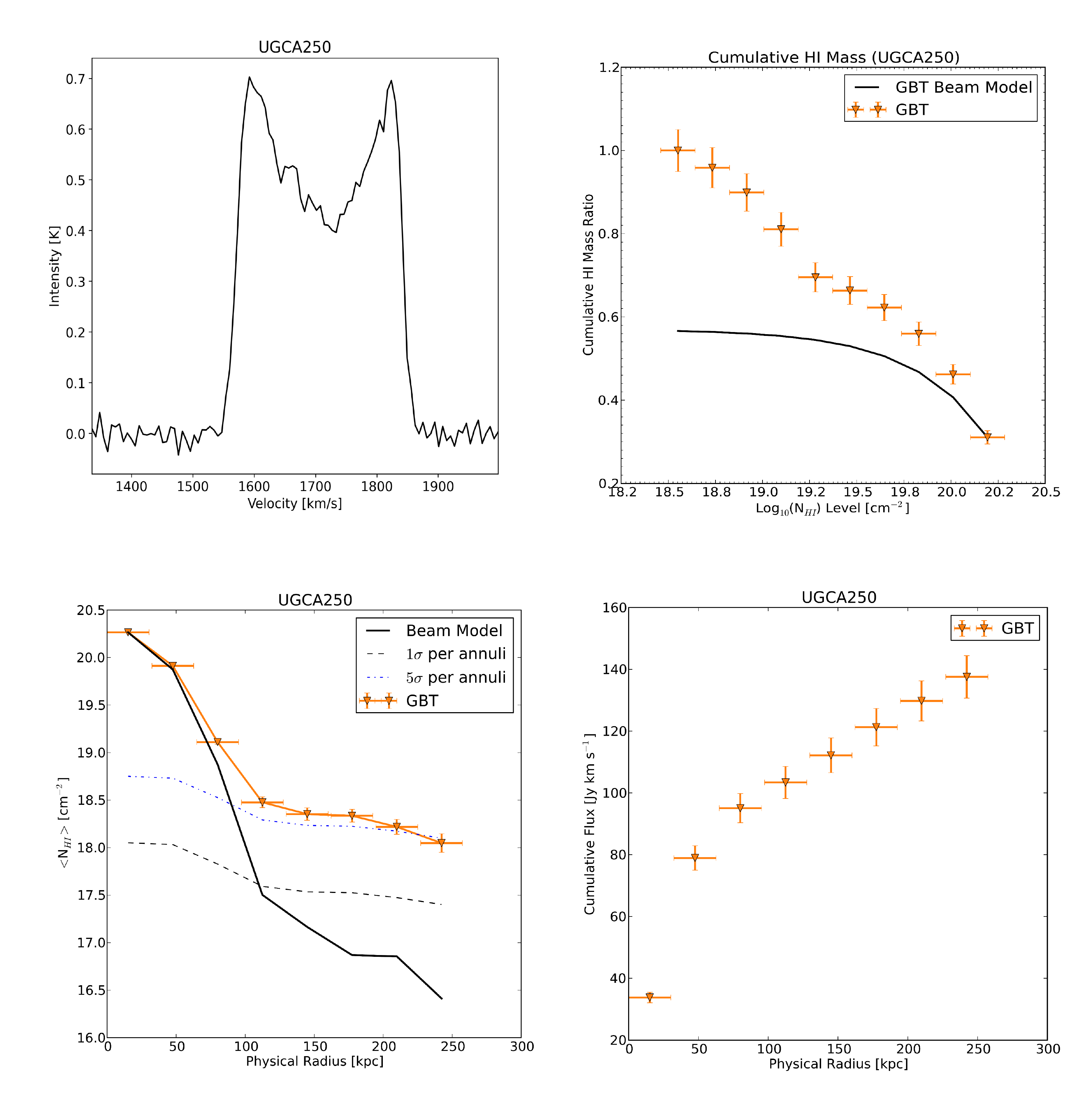}
\caption{Same as Figure \ref{fig:chapter3:NGC7424_4x4} for UGCA250. 
\label{fig:chapter3:UGCA250}}
\clearpage
\end{figure*}

\end{document}